\documentclass[a4paper,11pt]{article}
\pdfoutput=1
\usepackage{jheppub}
\usepackage{amsmath}
\usepackage{float}
\usepackage{dirtytalk}
\usepackage{bibentry}
\usepackage{comment}

\newcommand{\Tr}{\mathrm{Tr}\,}

\title{\boldmath Recursion relations for scattering amplitudes with massive particles }
\author{Sourav Ballav}
\affiliation{Institute of Mathematical Sciences \\
Homi Bhabha National Institute (HBNI)\\
IV Cross Road, C.~I.~T.~Campus, \\
  Taramani, Chennai, 600113  Tamil Nadu, India \\}
\emailAdd{sballav@imsc.res.in}

\author{ and Arkajyoti Manna}
\emailAdd{arkajyotim@imsc.res.in}

\abstract{We use the recently developed massive spinor-helicity formalism \citep{Arkani-Hamed:2017jhn} of Arkani-Hamed et al. to study a new class of recursion relations for tree-level amplitudes in gauge theories. These relations are based on a combined complex deformation of massless as well as massive external momenta. We use these relations to study tree-level amplitudes in scalar QCD as well as amplitudes involving massive vector bosons  in the Higgsed phase of Yang-Mills theory.  We prove the validity of our proposal by showing that in the limit of infinite momenta of two of the external particles, the amplitude once again is controlled by an enhanced Spin-Lorentz symmetry paralleling the proof of BCFW shift for massless gauge theories. Simple examples illustrate that the proposed shift may lead to an efficient computation of tree-level amplitudes.}


\begin{document}
\maketitle
\flushbottom

\section{Introduction and Summary }

The S-matrix program of quantum field theory has witnessed a number of remarkable developments in past three decades. 
On one hand, the study of analytic structure of S-matrix continues to  reveal strikingly new insights with potential to revolutionise the entire edifice of Quantum field theory \citep{ArkaniHamed:2010gg, Arkani-Hamed:2013jha, Cachazo:2013gna, Cachazo:2013iea} and on the other hand, on-shell techniques like Britto-Cachazo-Feng-Witten (BCFW) \citep{Britto:2004ap,Britto:2005fq} recursion relations and generalised unitarity \citep{Bern:2011qt} have made seemingly impossible computations possible within a few pages. The latter developments are directly responsible for the NLO revolution in QCD and have even been used in computing  classical observable quantities such as potential in the Binary black-holes up to high order in Post Newtonian and Post Minkowskian expansion \citep{Bern:2019crd}.

The basic idea of the recursion relations is to obtain higher-point amplitudes in terms of lower-point (or more precisely three-point on-shell) amplitudes. In contrast to the recursion relation obtained from unitarity condition, BCFW recursion operates solely on on-shell amplitudes and does so by working in the space of complex external momenta. The desired amplitude is a residue of a meromorphic function with a simple pole at infinity. For the class of quantum field theories with a specific high momentum behaviour, recursion relations completely determine the tree-level on-shell S-matrix from three-point amplitudes. There are several variants of recursion relations which differ by the number of external momenta that one analytically continues in the complex domain. By far the most widely used and efficient version is the BCFW recursion relation in which precisely two of the external momenta are complexified such that momentum conservation and on-shell conditions are preserved.  We refer the reader to \citep{elvang_huang_2015} for the most succinct introduction to these ideas. 

Scattering amplitudes are functions on the kinematic space of (generalised) Mandelstam invariants of external momenta. 
However this space is rather complicated due to non-linear constraints like the Gram determinant conditions and hence choice of appropriate variables to co-ordinatise the kinematic space becomes paramount in efficient computations of the scattering amplitudes. In Four dimensions, when external particles are massless, the most well known choice is so-called spinor-helicity variables. Using spinor-helicity variables and implementing BCFW recursion relations in terms of spinor-helicity variables  render computations of tree-level amplitudes in gauge theories and gravity strikingly simple as opposed to Feynman diagrams. However until recently things were less clear if some of the external states  were massive. Although spinor formalism for massive particles has been known for decades, these variables were not the most convenient for computation of S-matrix as they were not little group covariant \citep{Kleiss:1985yh, Schwinn:2005pi, Schwinn:2006ca, Badger:2005jv, Ozeren:2006ft}. However recently Arkani-Hamed et al.\citep{Arkani-Hamed:2017jhn} have introduced a little group covariant spinor-helicity formalism for massive particles. This formalism also allows one to take suitable massless limit of the massive amplitude in a systematic manner. We  use this formalism throughout this paper in computing massive scattering amplitude.

However even in the absence of a spinor-helicity formulation for massive particles, tree amplitudes involving a set of  massive and at least two massless particles have been analysed  by using the BCFW recursion\citep{Badger:2005jv, Badger:2005zh, Ochirov:2018uyq}. For instance in \citep{ Badger:2005zh} several lower-point tree-level amplitudes with  massive scalars and gluons were computed using the conventional BCFW relations. In \citep{Badger:2005jv} this was extended to computations of scattering amplitude involving vector bosons (spin-1) and fermions scattering via gluons. In a notable work \citep{Schwinn:2007ee}, the authors derived on-shell recursion relations for all Born amplitudes in QCD and gave a compact expression for amplitude involving a pair of massive quarks and arbitrary number of gluons. Recently in a beautiful paper, Ochirov \citep{Ochirov:2018uyq} generalised this computation by using the newly developed massive spinor-helicity formalism of \citep{Arkani-Hamed:2017jhn}. In particular using the BCFW shift on pair of massless external particles (gluons), Ochirov computed four-point amplitude and then proposed  formulae for two specific  $n$-point amplitudes consistent with the results derived previously \citep{Schwinn:2007ee, Ferrario:2006np}. And finally, in \citep{Schwinn:2005pi, Franken:2019wqr, Falkowski:2020aso} scattering amplitudes involving massive particles have been first studied using multi-line complex shifts. However in these works, the massive momenta are decomposed in terms of certain null vectors and the computation is brought closer to that of amplitude involving only massless particles. Recently, in \citep{Herderschee:2019ofc, Herderschee:2019dmc}  recursion relation for massive supersymmetric amplitudes have been developed using a massive super-BCFW shift.

In a certain sense  \cite{Arkani-Hamed:2017jhn} has unified the kinematic space of (massive and massless) external momenta into spinor-helicity basis and hence a natural question to ask is, how democratic is the BCFW (and in general recursion) technique of complexifying external momenta. That is, is it possible to derive ``BCFW type" recursion relations by complexifying massive instead of massless external states. Are these recursion relations more efficient than the ones derived earlier for similar external configurations? Under what conditions are these shifts valid? 
  
 In this paper we study these questions by proposing a generalisation of BCFW recursion to the case where one massive and one massless momenta are complexified. There has been earlier work in this direction in \citep{Aoude:2019tzn} in which a particular massive-massless shift was proposed and it was used to compute the scattering amplitude involving two massive vector bosons and two photons. We extend this by giving a classification of all possible massive-massless shifts and also discuss in detail the validity of the various shifts. In particular we show that not all possible shifts lead to a valid recursion relation. 
 The shift of external configurations, (that is complexification of real momenta by adding complex deformations)  allows us to work with the formalism of \citep{Arkani-Hamed:2017jhn} seamlessly.  
 We analyse the amplitudes in scalar QCD and Higgsed Yang-Mills theory for computing massive scalar and massive vector boson amplitudes respectively\cite{Badger:2005jv, Franken:2019wqr}. We show that the recursion relations lead to new compact formulae for five-point amplitudes involving massive vector bosons and gluons.

As mentioned previously we provide  a proof for the validity of the massive-massless shift. A comprehensive study of the validity of original BCFW shift for different theories has previously appeared in the seminal work by Arkani-Hamed et. al. \citep{Nima2008}. Our proof for massive-massless shift follows similar line of work but is more involved due to presence of an extra longitudinal mode corresponding to the massive particle.  For completeness, we also extend the proof of \cite{Nima2008} to the case of massive scattering amplitudes and prove the validity of  the traditional massless-massless shift for these theories.

This paper is structured as follows. We start with a brief review of spinor-helicity formalism for massive particles and structure of three-point amplitude \citep{Arkani-Hamed:2017jhn} in section  \ref{section:2}. In section \ref{section:b}, we  discuss the massive-massless shift and BCFW recursion scheme for amplitudes when massive particles are involved. In section \ref{section:4} a consistent deformation of spinor-helicity variables has been proposed to achieve the desired shift. Using this proposal, few four- and five-particle amplitude have been calculated. Section \ref{section:5} and \ref{section:6} contain the proof for validity of the massive-massless and massless-massless shifts respectively. We conclude our present work and mention some possible future direction in section \ref{section:7}. In appendix \ref{appendix: a} we give some technical details of deformed massive and massless polarization. In appendix \ref{appendix:B} we discuss a class of invalid massive-massless shifts.

\section{Review of Massive Spinor-Helicity Formalism}\label{section:2}

The scattering amplitude is defined as an inner product of the \say{in} and \say{out} states. These are one-particle states, labelled by momentum (and other discrete quantum numbers) and the little group indices of the particles. Thus the amplitude naturally carries little group indices. Let us consider a scattering amplitude involving helicity $h$ massless particles and a massive spin $S$ particle in four dimensions. The little group for massless particles is the group of transformations that map the two dimensional plane to itself, which is $ISO(2)$. Under its action the amplitude scales as $t^{-2h}$ where $t$ is a non-zero complex number. The little group associated to the massive particle is $SU(2)$. We use the symmetric $2S$ representation of the $SU(2)$ group to label the little group indices of the massive particle, following \citep{Arkani-Hamed:2017jhn}. The amplitude involving massless particles of momentum $p_j$ and massive particles of momentum $p_i$ therefore transforms in the following manner under the little group transformation:
\begin{align}
\mathcal{A}^{h}_{I_1I_2...I_{2S}}(p_i,p_j)\longrightarrow t_j^{-2h}W_{I_1}~^{J_1}W_{I_2}~^{J_2} \cdots W_{I_{2S}}~^{J_{2S}}\mathcal{A}^{h}_{J_1J_2\ldots J_{2S}}(p_i,p_j)~.
\end{align}  
Here $W$'s are $SU(2)$ matrices and $\mathcal{A}^{h}_{I_1I_2...I_{2S}}(p_i,p_j)$ is referred to as the natural amplitude in \citep{Arkani-Hamed:2017jhn}. The above transformation rule implies that the amplitude is symmetric under the permutation of little group indices $(I_1,..,I_{2S})$ and $(J_1,..,J_{2S})$ for the massive particles. In order to make manifest the little group transformation properties of the amplitude we now introduce the spinor-helicity variables.

 We are interested in computing scattering amplitudes in four dimensions, so we briefly review the four dimensional spinor-helicity formalism. The basic goal of this formalism is to express on-shell momentum in terms of spinor variables that transform in the $\left(0,\frac{1}{2}\right)$ and $\left(\frac{1}{2},0\right)$ representation of the Lorentz group. These spinors are called spinor-helicity variables. 
 A momentum $4$-vector $p^\mu$ can be expressed as a $2\times 2$ hermitian matrix $p_{\alpha \dot{\alpha}}=p_\mu \sigma ^\mu_{\alpha \dot{\alpha}}$, where $\sigma ^0_{\alpha \dot{\alpha}}$ is identity matrix and $\vec{\sigma }_{\alpha \dot{\alpha}}$ are Pauli matrices. The determinant of the matrix $p_{\alpha \dot{\alpha}}$ is $p_{\mu}p^{\mu}=p^2$ which is zero for a massless particle. So the momentum for a massless particle ($p_j$) can be written in terms of two 2-component spinors $(\lambda _{j\alpha},\tilde{\lambda}_{j\dot{\alpha}})$ as
\begin{align}
p_{j\alpha \dot{\alpha}}:=\lambda _{j\alpha}\tilde{\lambda}_{j\dot{\alpha}}\equiv|\lambda_j\rangle [\tilde{\lambda}_j|~.  
\end{align}
Here $(\lambda _{j\alpha }, \tilde{\lambda}_{j\dot{\alpha}})$ are the spinor-helicity variables. 

The matrix $p_{\alpha \dot{\alpha}}$ for a massive particle ($p_i$) has non-zero determinant. Therefore it can be expressed as a linear combination of rank one matrices \citep{Arkani-Hamed:2017jhn}:
\begin{align}
p_{i\alpha \dot{\alpha}}=\sum _{I=1}^2 \lambda ^I_{i\alpha}\tilde{\lambda}_{i\dot{\alpha}I}\equiv \sum _{I=1}^2 |\lambda ^I_i\rangle [\tilde{\lambda}_{iI}| ~.
\end{align}
The massive spinor-helicity variables $(\lambda _{i\alpha I}, \tilde{\lambda}_{i\dot{\alpha}J})$ carry an extra set of indices $(I,J)$ which labels the little group $SU(2)$. 
These can be expanded in terms of a pair of massless spinor-helicity variables $(\lambda _\alpha, \tilde{\lambda}_{\dot{\alpha}})$ and  $(\eta _\alpha, \tilde{\eta}_{\dot\alpha})$ \citep{Arkani-Hamed:2017jhn} as follows: 
\begin{align}
	\lambda _{iI}^\alpha &=\lambda _i ^\alpha \xi ^+_I-\eta _i^\alpha \xi ^-_I \label{hevar1}\\
	\tilde{\lambda}_{iJ}^{\dot{\alpha}}&= -\tilde{\lambda}_i^{\dot{\alpha}} \xi ^-_J+\tilde{\eta}_i ^{\dot{\alpha}}\xi ^+_J~, \label{hevar2}
\end{align}
where  $\xi ^{\pm}_I$ are two orthonormal 2-component vectors which serve as a basis in the $SU(2)$ space. This expansion allows one to take the high energy limit of the massive spinor-helicity variables which in turn will prove to be useful while analyzing the high energy limit of scattering amplitudes involving massive particles \citep{Arkani-Hamed:2017jhn}. 
Let us see how this works in practice at the level of the massive spinor-helicity variables. If we write the massive four momentum as 
$$
p_i ^\mu \equiv (p^0_i,|\vec{p}_i|\sin(\theta)\cos(\phi),|\vec{p}_i|\sin(\theta)\sin(\phi),|\vec{p}_i|\cos(\theta))~,
$$ 
then the massless spinors can be expressed in the limit $p^0_i >>|\vec{p}_i|$ as
\begin{align}
\lambda _{i\alpha}\rightarrow \sqrt{2p^0_i} \begin{pmatrix}
\cos (\theta/2)\\
\sin (\theta /2)e^{i\phi}
\end{pmatrix}~,
\quad \eta _{i\alpha}\rightarrow \frac{m_i}{\sqrt{2p^0_i}} \begin{pmatrix}
-\sin (\theta /2)e^{-i\phi}\\
\cos(\theta /2)
\end{pmatrix}~.
\end{align}
This shows that $\eta _{i\alpha}$ is proportional to the mass $m_i$ and vanishes in this limit. A similar result holds for $\tilde{\eta}_{i\dot{\alpha}}$. So in the high energy limit both $(\eta _{i\alpha},\tilde{\eta}_{i\dot{\alpha}})$ vanish.

Next we move on to defining polarization tensors for massive and massless particles. For massless particles, this is well studied and the polarization vector is expressed in terms of spinor-helicity variables in the following way \citep{elvang_huang_2015}:
\begin{align}
e^\mu _+(j) =\frac{\langle \xi |\sigma ^\mu |\tilde{\lambda}_j]}{\langle \xi {{\lambda}}_j\rangle}~, \qquad e^\mu _-(j) =\frac{\langle \lambda _j |\sigma ^\mu |\tilde{\eta}]}{[\tilde{\lambda}_j\tilde{\eta}]}~.
\end{align}
Here $\xi ^\alpha$ and $\tilde{\eta}^{\dot{\alpha}}$ are reference spinors. The polarization tensor associated to a massive particle transforms as a  bi-fundamental under the Lorentz and $SU(2)$ little group transformations. For a massive spin-1 particle the polarization tensor is given in terms of the massive spinor-helicity variables as follows \citep{Guevara:2018wpp, Chung:2018kqs}: 
\begin{align}
{e}^\mu _{IJ}(i)=\frac{1}{2\sqrt{2}m}\langle {\lambda} _{i(I}|\sigma ^\mu |{\tilde{\lambda}}_{iJ)}]. \label{massivepol}
\end{align}

\subsection{Three-particle amplitude: basic building blocks}\label{section: minimal}

The three-particle amplitude for massless particles is highly constrained due to momentum conservation. It is only a function of either $\lambda _\alpha$ or $\tilde{\lambda}_{\dot{\alpha}}$ and the amplitude is completely fixed by its scaling property under the little group action. For three massless particles of helicity $(h_1,h_2,h_3)$, the scattering amplitude is:
\begin{equation}
\mathcal{A}_3^{h_1h_2h_3}[1,2,3]=
\begin{cases} 
g [12]^{h_1+h_2-h_3}[23]^{h_2+h_3-h_1}[31]^{h_3+h_1-h_2} &  \text{for $h_1+h_2+h_3 >0$}\\
 g'\langle 12\rangle ^{h_3-h_1-h_2} \langle 23\rangle ^{ h_1-h_2-h_3} \langle 31\rangle ^{h_1-h_2-h_3} & 
\text{for $h_1+h_2+h_3 <0$}~.
\end{cases}
\end{equation} 
The conditions on the sum of helicities ensure that the amplitude has a smooth vanishing limit for real momenta in Minkowski signature (as individual spinor products vanish in this signature). 

Next we will consider three-particle amplitudes involving two massive particles with same mass $m$ and spin $S$ and a massless particle. Unlike the massless case, this three-particle amplitude is not fully determined by the little group action. We will be interested only in the minimally coupled massive and massless particles. In this case the three-particle amplitudes are given by \citep{Arkani-Hamed:2017jhn}: 
\begin{align}
\mathcal{A}_3^{+h}(\textbf{1},\textbf{2},3^h)=gx_{12}^h \frac{\langle \textbf{1}\textbf{2}\rangle ^{2S}}{m^{2S-1}}~,\quad \mathcal{A}_3^{-h}(\textbf{1},\textbf{2},3^{-h})=gx_{12}^{-h} \frac{[ \textbf{1}\textbf{2}] ^{2S}}{m^{2S-1}}~.\label{minimalthreepoint}
\end{align}
Here $x_{12}$ is a non local factor which arises due to the presence of massive particles with identical masses. It is  defined as 
\begin{align}
x_{12}=\frac{\langle \zeta |p_1|3]}{m\langle \zeta 3\rangle} \quad \text{or}\quad x_{12}^{-1}=\frac{\langle 3|p_1|\zeta]}{m[3\zeta]}~,
\end{align} 
where $\zeta ^\alpha$ is a reference spinor. We denote the massive spinor-helicity associated to massive particles using a bold notation. For convenience we have omitted the little group indices for massive particles in the amplitude, but they are captured in the definition of the bold face spinor products. These spinor products are defined as a symmetric combination of normal spinor products carrying $SU(2)$ indices. For example 
\begin{align}
\langle \textbf{1}\textbf{2}\rangle ^2&:= \langle 1^{I_1}2^{J_1}\rangle \langle 1^{I_2}2^{J_2}\rangle +\langle 1^{I_2}2^{J_1}\rangle \langle 1^{I_1}2^{J_2}\rangle ~;\\
\langle 3\textbf{2}\rangle ^2 &:=\langle 32^{J_1}\rangle \langle 32^{J_2}\rangle~. 
\end{align}

\section{BCFW Recursion Scheme for Massive Particles}\label{section:b}

In this section we review the generalization of the BCFW recursion relation for (tree-level) scattering amplitudes involving both massive and massless particles \citep{Britto:2005fq, Badger:2005zh, Franken:2019wqr}. In essence the idea of BCFW recursion is to derive an $n$-point scattering amplitude in terms of lower-point on-shell amplitudes. At the heart of this approach is the unitarity of quantum field theories which implies that tree-level amplitude factorises into lower-point sub-amplitudes when a propagator goes on-shell. 

We are interested in tree amplitudes with particle configurations such that  there is atleast one massless particle. In order to derive the on-shell recursion relation, two external momenta are shifted by null momenta such that  momentum conservation and on-shell conditions for the shifted momenta are still obeyed. Consider two external momenta $(p_i, p_j)$ which are analytically continued to the complex plane with lightlike momentum $r^\mu$ in the following way
\begin{align}
p_i^\mu \rightarrow \widehat{p}_i^\mu= p_i^\mu-zr^\mu ~;\qquad p_j^\mu \rightarrow \widehat{p}_j^\mu= p_j^\mu+zr^\mu . \label{shift}
\end{align}
Here $z$ is a deformation parameter. The on-shell property is maintained by imposing the following constraints
\begin{align}
	p_i\cdot r=p_j\cdot r =0.
\end{align}
The scattering amplitude $\mathcal{A}_n$ with undeformed momenta can be related to the deformed amplitude $\widehat{\mathcal{A}}_n(z)$ by using Cauchy's theorem 
\begin{align}
\mathcal{A}_n=\widehat{\mathcal{A}}_n(0)=\frac{1}{2\pi i}\oint _{\Gamma _0} \frac{\widehat{\mathcal{A}}_n(z)}{z}dz=-\sum _{z_I} \text{Res}\left(\frac{\widehat{\mathcal{A}}_n(z)}{z}\right)_{z=z_I} +\mathcal{R}_n .\label{complexint}
\end{align}
The contour $\Gamma _0$ encloses the pole at the origin. $\mathcal{R}_n$ is the boundary term that comes from the contribution of the pole at infinity. All other simple poles of the amplitude are denoted by $z_I$. 

In the case of tree-level scattering amplitudes, the poles in the $z$-plane are correlated with the poles in kinematic space. The deformed amplitude $\widehat{\mathcal{A}}_n(z)$ has simple poles in the kinematic space whenever an internal propagator of the form $\frac{1}{\widehat{P}^2-m^2}$ goes on-shell. But when this occurs the amplitude $\widehat{\mathcal{A}}_n(z)$ factorizes into two lower-point on-shell sub-amplitudes. Hence we have 
\begin{align}
\widehat{\mathcal{A}}_n(z)=\sum _{I}\widehat{\mathcal{A}}_{l+1}\frac{1}{\widehat{P}_I^2-m^2}\widehat{\mathcal{A}}_{r+1}~,
\end{align}
where $I$ corresponds to different scattering channels and $n=l+r$. It is important to note that the sub-amplitudes are functions of shifted momenta. To get the required pole structure, we simply write the shifted propagator in terms of the propagator with unshifted momenta:
\begin{align}
\widehat{P}_I^2|_{z_I}=m^2\Rightarrow (P_I+z_I\,p_j)^2=m^2 \Longrightarrow            \frac{1}{\widehat{P}_I^2-m^2}=-\frac{z_I}{z-z_I}\frac{1}{{P_I}^2-m^2}.
\end{align}
Now let us focus on the boundary term $\mathcal{R}_n$ at $z \rightarrow \infty$. This term can not be computed from {\bf a single} recursion relation \citep{Britto:2004ap, Britto:2005fq}. So a BCFW shift such as the one in  \eqref{shift} is called valid when this  boundary term vanishes. Note that $\mathcal{R}_n$ is vanishing when we have
\begin{align}
\widehat{\mathcal{A}}_n(z) \rightarrow 0 ~,\qquad \text{as} ~z\rightarrow \infty .\label{validity}
\end{align}
So we treat \eqref{validity} as a condition for a valid BCFW shift. Finally let us give the BCFW recursion scheme to compute the amplitude  
 \begin{align}
 \mathcal{A}_n=-Res\left(\frac{1}{z}\sum _{I} \widehat{\mathcal{A}}_{l+1}(z) \frac{1}{\widehat{P}_I^2-m^2}\widehat{\mathcal{A}}_{r+1} (z)\right)\Bigg|_{z_I}=\sum _{I} \widehat{\mathcal{A}}_{l+1}(z_I) \frac{1}{P_I^2-m^2}\widehat{\mathcal{A}}_{r+1} (z_I).\label{bcfwrec}
 \end{align}
\begin{figure}
\includegraphics[scale=.35]{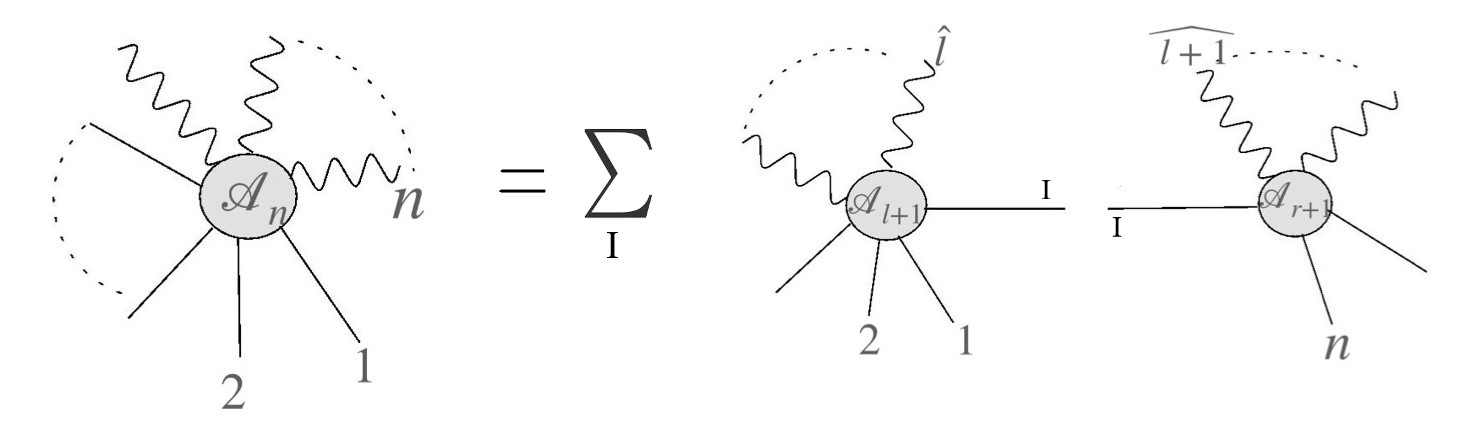}
\caption{BCFW recursion scheme}
\end{figure} 

We recall that the only those diagrams in which the two deformed momenta  are on opposite sides of the on-shell propagator contribute to the residue at $z\, =\, 0$, resulting in enormous computational  simplifications as compared to the computation of Feynman diagrams. 

\section{Proposal for a Valid Massive-Massless Shift }\label{section:4}

Now we consider a two-line BCFW shift involving a massive particle with momentum $p_i$ and a massless particle with momentum $p_j$ and positive helicity satisfying the same shift equation as in \eqref{shift}. To achieve this massive-massless shift, we propose the following deformation of massless and massive spinor-helicity variables: 
\begin{align}
\text{massive shift }:& \qquad	\widehat{\lambda}^I_{i\alpha}=\lambda ^I_{i\alpha}  ~,\qquad \widehat{\tilde{\lambda}}^I_{i\dot{\alpha}}=\tilde{\lambda} _{i\dot{\alpha}}^I-\frac{z}{m}\tilde{\lambda}_{j\dot{\alpha}}[i^Ij]~, 
	\label{massiveshift}\\
\text{massless shift }: & \qquad    \widehat{\tilde{\lambda}}_{j\dot{\alpha}} =\tilde{\lambda}_{j\dot{\alpha}}~, \qquad \widehat{\lambda}_{j\alpha}=\lambda _{j\alpha} +\frac{z}{m}p_{i\alpha \dot{\beta}}\tilde{\lambda}_j^{\dot{\beta}}~,\label{masslessshift}
\end{align}
where we have chosen the null momentum $r_{\alpha \dot\alpha} = r_{\mu}\sigma^{\mu}_{\alpha\dot\alpha}$ to be:
\begin{equation}
\label{refrdef}
r_{\alpha \dot{\alpha}} =\frac{p_{i\alpha \dot{\beta}}}{m}\tilde{\lambda}_j^{\dot{\beta}} \tilde{\lambda}_{j\dot{\alpha}}~.
\end{equation}
We denote such a shift as $[\textbf{i}j\rangle $ of the type $[\textbf{m}+\rangle$. Note that under the little group transformation both shifted spinor-helicity variables transform covariantly. Hence we can directly use the shifted variables in the amplitudes written in a manifestly little group covariant form \citep{Arkani-Hamed:2017jhn}.  
We also note that the deformed massless polarization vector for a particular choice of the reference spinor is given by
\begin{align}
\widehat{e}^\mu _+(j) =\frac{\sqrt{2}m r^\mu}{\langle \lambda _j |p_i|\tilde{\lambda}_j]}=\kappa r^\mu ~.\label{map}
\end{align}
We refer to appendix \ref{appendix:c} for more details. 

We use the shifted spinor-helicity variables and the associated polarization vectors in the proposed generalized recursion relation:
\begin{align}
\mathcal{A}_n=\sum _{I} \widehat{\mathcal{A}}_{l+1}(z_I) \frac{1}{P^2-m^2}\widehat{\mathcal{A}}_{r+1} (z_I)+\sum _{J} \widehat{\mathcal{\tilde{A}}}_{l+1}(z_J) \frac{1}{P^2}\widehat{\mathcal{\tilde{A}}}_{r+1} (z_J)~.\label{BCFWrecursion}
\end{align}
We use the recursion relation to compute several four- and five-point amplitudes in two models: i) scalar QCD with massive scalars scattering off gluons and ii) spontaneously broken non-abelian gauge theory whose spectrum consists of gluons and massive vector bosons. The two terms in the recursion represent contributions to the amplitude with massive and massless intermediate propagators respectively. We will consider only colour-ordered amplitudes instead of fully colour dressed tree amplitudes as the latter can be constructed from the former using the well known colour decomposition \citep{DelDuca:1999rs, Dixon:1996wi, Johansson:2015oia, Ochirov:2019mtf, Maltoni:2002mq, Melia:2015ika}.

\subsection{ Scalar QCD: Compton amplitude }\label{H}
Let us consider a $2\rightarrow 2$ scattering involving two massive scalars with momentum $(p_1,p_4)$ and two  positive helicity gluons with momentum $(p_2,p_3)$.  Using the conventions in \eqref{massiveshift} and  \eqref{masslessshift}, we consider the $[\textbf{1}2\rangle $ shift on the massive momentum  $p_1$ and the massless momentum $p_2$: 
\begin{align}
\widehat{p}_1^\mu=p_1^\mu-zr^\mu~,\quad \widehat{p}_2^\mu=p_2^\mu +zr^\mu~.
\end{align}
In this example there exists no $4$-point scalar-gluon contact term, so the four-particle amplitude can be constructed via three-point amplitudes using the recursion \eqref{BCFWrecursion}. 

\begin{figure}[H]
\includegraphics[scale=.35]{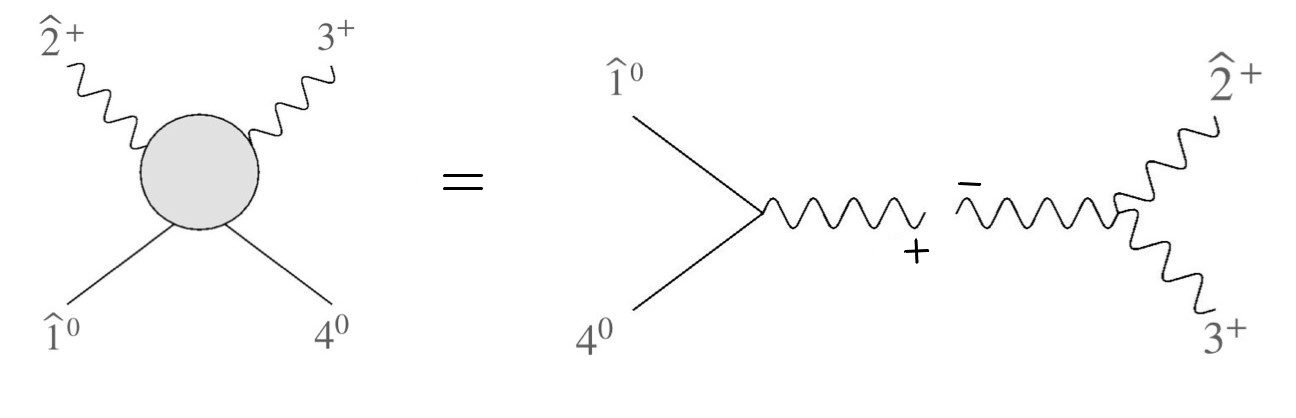}
\caption{Four-particle amplitude in scalar QCD}
\label{4ptscalarQCD}
\end{figure}
The amplitude can be constructed from left and right sub-amplitudes in the Figure \ref{4ptscalarQCD} as follows: 
\begin{align}
\mathcal{A}_4[\textbf{1}^0,2^+,3^+,\textbf{4}^0]=mg^2 \widehat{x}_{14}\frac{1}{s_{23}} \frac{[23]^3}{[2\widehat{I}][3\widehat{I}]}=m^2g^2\frac{[23]^3}{\langle \widehat{I}|p_4|3] [\widehat{I}2]s_{23}}~,
\end{align}
where g is a dimensionless coupling and the non-local $x$-factor given by
\begin{align}
\widehat{x}_{14}=m\frac{[\widehat{I}3]}{\langle \widehat{I}|p_4|3]}~.
\end{align}
We denote the standard Mandelstam variables as $s_{mn}=(p_m+p_n)^2$. Now the $\widehat{I}$-dependent terms can be simplified by using
\begin{align}
\langle \widehat{I}|p_4|3] [\widehat{I}2]=-[2|p_1\cdot p_4|3]+m^2[23]~.
\end{align}
In the last step we have used $[2\widehat{1}^I]=[21^I]$. Using four-particle kinematics we can write
\begin{align}
[2|p_1\cdot p_4|3]=(s_{12}-m^2)[23]+m^2[23]=[23]s_{12}
\end{align}
So the four-particle amplitude becomes
\begin{align}
\mathcal{A}_4[\textbf{1}^0,2^+,3^+,\textbf{4}^0]=-m^2g^2\frac{[23]}{\langle 23\rangle (s_{12}-m^2)}
\end{align}
This matches the result given in \citep{Badger:2005zh}. The amplitude involving opposite helicity gluons $(2^+,3^-)$ can be similarly computed using the BCFW shifts $\,$\eqref{massiveshift} and \eqref{masslessshift}:
\begin{align}
\mathcal{A}_4[\textbf{1}^0,2^+,3^-,\textbf{4}^0]=g^2\frac{\langle 3|p_1|2]^2}{s_{23}(s_{12}-m^2)}~. \label{scalarqcd4}
\end{align}
This result agrees with the Compton amplitude given in \citep{Arkani-Hamed:2017jhn}.
\subsection{Scalar QCD : five-particle amplitude}

We now compute colour-ordered five-particle amplitude corresponding to the scattering of two massive scalar particles and three gluons with arbitrary helicity. We will use the $[\textbf{2}3\rangle$ shift  to compute this amplitude. The five-point amplitude can be obtained from three- and four-point amplitudes via the following recursion relation:
\begin{align}
\mathcal{A}_5[\textbf{1}^0,\textbf{2}^0,3^{h_1},4^{h_2},5^{h_3}]=\sum _{I} \widehat{\mathcal{A}}_{3}(z_I) \frac{1}{P^2-m^2}\widehat{\mathcal{A}}_{4} (z_I)+\sum _{J} \widehat{\mathcal{\tilde{A}}}_{3}(z_J) \frac{1}{P^2}\widehat{\mathcal{\tilde{A}}}_{4} (z_J)~.\label{scalarfivepoint}
\end{align}
 Due to the choice of this shift, there will be only two colour-ordered diagrams. Apart from these two diagrams there are other diagrams involving a massive internal propagator, but they vanish as a massive spin $s$ particle cannot decay into two helicity-$h$ massless particles when $s<2|h|$. We now specialize to the case in which the gluons have  the following helicity configuration:
\begin{figure}[H]
\includegraphics[scale=.3]{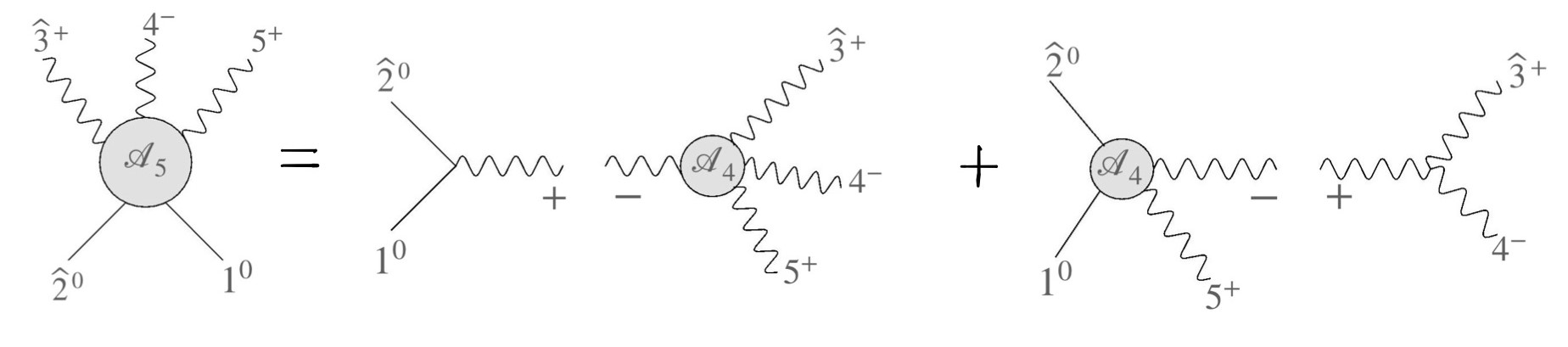}
\caption{Five-particle amplitude in scalar QCD} \label{figure:3}
\end{figure} 
%
%
\noindent
The full colour-ordered amplitude can be computed by summing over the two diagrams shown in the R.H.S of figure \eqref{figure:3}.\\
\textbf{First diagram :} \\
Using eqn \eqref{scalarfivepoint} the contribution from the first diagram is
\begin{align}
\mathcal{A}_5^{(I)}\left[\textbf{1}^0,\textbf{2}^0,3^+,4^-,5^+\right]=\widehat{\mathcal{A}}_3\left[ \textbf{1}^0,\widehat{\textbf{2}}^0,\widehat{I}^+\right]\frac{1}{s_{12}}\widehat{\mathcal{A}}_4\left[\widehat{I}^-,\widehat{3}^+,4^-,5^+\right].
\end{align}
We consider only minimal coupling between the scalar and the gluon. Then the three-particle amplitude follows from \eqref{minimalthreepoint} with $\widehat{x}_{12}=m\frac{[5\widehat{I}]}{\langle \widehat{I}|p_1|5]}$, while the massless four-particle amplitude is given by the Parke-Taylor formula \citep{PhysRevLett.56.2459}:
\begin{align}
 \label{fivethreescalar1}
\widehat{\mathcal{A}}_3\left[ \textbf{1}^0,\widehat{\textbf{2}}^0,\widehat{I}^+\right]&=gm^2 \frac{[5\widehat{I}]}{\langle \widehat{I}|p_1|5]}~,\qquad 
\widehat{\mathcal{A}}_4\left[\widehat{I}^-,\widehat{3}^+,4^-,5^+\right]=\frac{[35]^4}{[\widehat{I}3][34][45][5\widehat{I}]}~.
\end{align}
Using \eqref{fivethreescalar1}, the contribution of first diagram is:
\begin{align}
\mathcal{A}_5^{(I)}\left[\textbf{1}^0,\textbf{2}^0,3^+,4^-,5^+\right]=m^2g^3 \frac{[35]^4}{s_{12}[34][45]\left([3|p_2\cdot p_1|5]+m^2[35]\right)}~. \label{scalarqcdfivepoint}
\end{align}
At first sight, the terms within parentheses in the denominator of the last expression may seem to lead to a spurious pole. But as we shall now show, this is not the case using a proof by contradiction. Suppose the term in parentheses vanishes. Then we have:
\begin{align}
[3|p_2\cdot p_1|5]&=\tilde{\lambda}_{3\dot{\alpha}}p^{\dot{\alpha}}_{2\beta}p^\beta _{1\dot{\gamma}}\tilde{\lambda}_5^{\dot{\gamma}}=-m^2 [35] \cr
\Rightarrow p^{\dot{\alpha}}_{2\beta}p^\beta _{1\dot{\gamma}}&=-m^2 \delta ^{\dot{\alpha}}_{\dot{\gamma}}~.
\end{align}
Thus this combination in the denominator vanishes only when the massive momenta $p_1$ and $p_2$ become collinear. This is of course impossible for massive particles and we have reached a contradiction. 

\textbf{Second diagram :}\\

Now let us move on to the second diagram. This diagram involves a four-particle amplitude involving two massive scalars and opposite helicity gluons, and a three gluon vertex. We have already computed the four-particle amplitude in earlier section. Using the recursion in \eqref{scalarqcdfivepoint}, we write the contribution of the second diagram as follows:
\begin{align}
\mathcal{A}_5^{(II)}\left[\textbf{1}^0,\textbf{2}^0,3^+,4^-,5^+\right]=\widehat{\mathcal{A}}_4\left[\textbf{1}^0,\textbf{2}^0,\widehat{I}^-,5^+\right] \frac{1}{s_{34}}\widehat{\mathcal{A}}_3\left[\widehat{I}^+,\widehat{3}^+,4^-\right]~.
\end{align}
Using the four-particle amplitude in \eqref{scalarqcd4} and simplifying the $\widehat{I}$ dependent terms the second diagram evaluates to
\begin{align}
\mathcal{A}_5^{(II)}\left[\textbf{1}^0,\textbf{2}^0,3^+,4^-,5^+\right]=g^3 \frac{\langle 54\rangle \langle 4|p_1|5]^2}{\langle 5\widehat{3}\rangle \langle 34\rangle \widehat{s}_{12}(s_{15}-m^2)}~.
\end{align}
The shifted spinor products are evaluated at the simple pole $\tilde{z}_I$, which originates from the internal massless propagator of the second diagram:
\begin{align}
(\widehat{p}_3+p_4)^2=0 \Rightarrow   \tilde{z}_I=\frac{m\langle 34\rangle}{\langle 4|p_2|3]}~. \label{x1}
\end{align}
Then the shifted spinor products can be expressed in terms of unshifted spinor-helicity variables as
\begin{align} 
\widehat{s}_{12}=s_{12}-\frac{\langle 34\rangle}{\langle 4|p_2|3]}[3|p_1\cdot p_2|3]~;\quad \langle 5\widehat{3}\rangle =\frac{\langle 54\rangle \langle 3|p_2|3]}{\langle 4|p_2|3]}~.
\end{align} 
Using all these expressions we obtain the contribution to the five-particle amplitude from the second diagram:
\begin{align}
\mathcal{A}_5^{(II)}\left[\textbf{1}^0,\textbf{2}^0,3^+,4^-,5^+\right]=g^3\frac{\langle 4|p_1|5]^2\langle 4|p_2|3]^2}{\langle 3|p_2|3]\langle 34\rangle (s_{15}-m^2)\left(\langle 4|p_2|3] s_{12}+{\langle 4|p_3\cdot p_1\cdot p_2|3]}\right)}~. 
\end{align}
Once again we can check that it is impossible for the term within parenthesis in the denominator to vanish by using similar arguments as before. Finally we obtain the colour-ordered five-particle amplitude as
\begin{multline}
\mathcal{A}_5 ^{\text{total}}\left[\textbf{1}^0,\textbf{2}^0,3^+,4^-,5^+\right]= g^3 m^2 \frac{[35]^4}{s_{12}[34][45]\left([3|p_2\cdot p_1|5]+m^2[35]\right)}  \\
+ g^3\frac{\langle 4|p_1|5]^2\langle 4|p_2|3]^2}{(s_{23}-m^2)\langle 34\rangle (s_{15}-m^2)\left(\langle 4|p_2|3] s_{12}+{\langle 4|p_3\cdot p_1\cdot p_2|3]}\right)}~. \label{scalarqcd5}
\end{multline}
The amplitude with exactly the same configuration of scattering particles has been computed in \citep{Badger:2005zh} using different methods. To make contact with their result we make use of the following identities: 
\begin{align}
[5|(p_3+p_4)\cdot p_2|3] 
&=-m^2[35]-[3|p_2\cdot p_1|5] ~,\\
 \langle 45\rangle \left([3|p_2\cdot p_1|5]+m^2[35] \right) &=-(\langle 4|p_3\cdot p_1\cdot p_2|3]+s_{12}\langle 4|p_2|3])
\end{align}
Using these two identities the five-particle amplitude can be re-expressed as (upto sign)
\begin{multline}
\mathcal{A}_5 ^{\text{total}}\left[\textbf{1}^0,\textbf{2}^0,3^+,4^-,5^+\right]= g^3 m^2 \frac{[35]^4}{s_{12}[34][45][5|(p_3+p_4)\cdot p_2|3]}  \\
- g^3\frac{\langle 4|p_1|5]^2\langle 4|p_2|3]^2}{(s_{23}-m^2)\langle 34\rangle \langle 45\rangle (s_{15}-m^2)[5|(p_3+p_4)\cdot p_2|3]}~,
\end{multline}
The above expression exactly matches the result given in \citep{Badger:2005zh}.

Similarly for all positive helicity gluons, the five-particle amplitude has the following expression
\begin{align}
\mathcal{A}_5[\textbf{1}^0,\textbf{2}^0,3^+,4^+,5^+]
=m^2g^3\frac{[5|(p_3+p_4)\cdot p_2|3]}{\langle 34\rangle \langle 45\rangle  (s_{23}-m^2) (s_{15}-m^2)}~.
\end{align}
which again agrees with the result in \citep{Badger:2005zh}.

\subsection{Compton amplitude with massive vector bosons}{\label{section4.3}}
Now let us focus on the scattering amplitude involving massive  spin-$1$ particles and gluons. Recall from section \ref{section: minimal}  that the 3-point amplitudes are given by the following expressions
\begin{align}
\mathcal{A}_3^{+1}(\textbf{1}^1,\textbf{2}^1,3^+)=gx_{12} \frac{\langle \textbf{1}\textbf{2}\rangle ^{2}}{m}~,\quad \mathcal{A}_3^{-1}(\textbf{1}^1,\textbf{2}^1,3^{-})=gx_{12}^{-1} \frac{[ \textbf{1}\textbf{2}] ^{2}}{m}~.
\end{align}
 We first compute the Compton amplitude with two massive spin-$1$ particles and two opposite helicity gluons.
\begin{center}
\includegraphics[scale=.3]{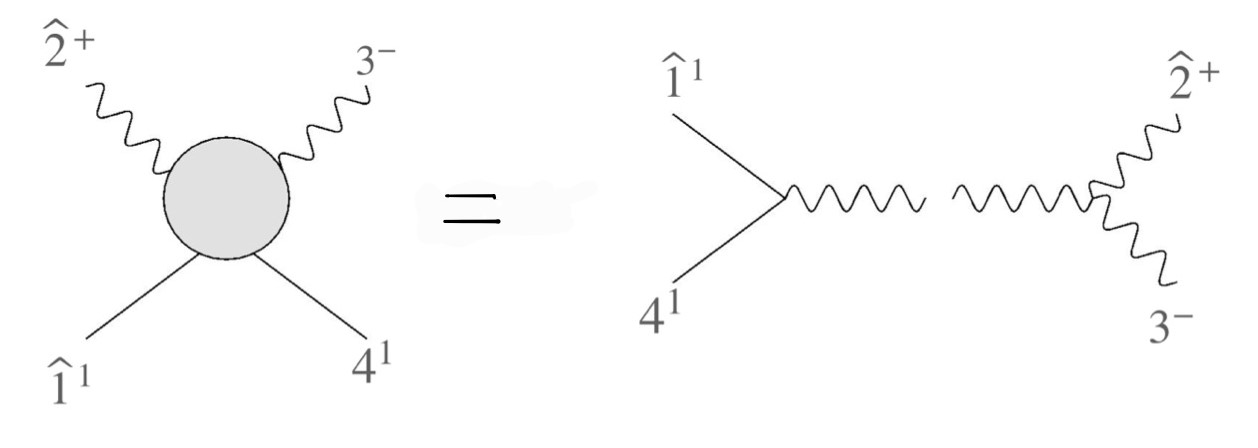}
\end{center}
Particles $(1,4)$ are massive spin-$1$ particles and the particles $(2,3)$ are gluons. We use $[\textbf{1}2\rangle $ BCFW shift to compute the only color-ordered amplitude from R.H.S of the above diagram. The relevant massive-massless shift is
\begin{align}
	\widehat{\lambda}_{2\alpha}=\lambda _{2\alpha} +\frac{z}{m}p_{1\alpha \dot{\beta}}\tilde{\lambda}_2^{\dot{\beta}} ~,\qquad \widehat{\tilde{\lambda}}_{2\dot{\alpha}} =\tilde{\lambda}_{2\dot{\alpha}}, \label{masslessshiftex1} 
\end{align}
and 
\begin{align}
	\widehat{\lambda}^I_{1\alpha}=\lambda ^I_{1\alpha}  ~;\qquad \widehat{\tilde{\lambda}}^I_{1\dot{\alpha}}=\tilde{\lambda} _{1\dot{\alpha}}^I-\frac{z}{m}\tilde{\lambda}_{2\dot{\alpha}}[1^I2].  \label{massiveshift1}
\end{align}
The four-particle amplitude can be constructed by using the BCFW recursion relation \eqref{BCFWrecursion}
\begin{align}
\mathcal{A}_{4}\left[\textbf{1},2^+,3^-,\textbf{4}\right] =\widehat{A}_3 \left[\widehat{\textbf{1}},\widehat{I}^-,\textbf{4}\right]\frac{1}{s_{23}}\widehat{A}_3\left[\widehat{I}^+,\widehat{2}^+,3^-\right] ~.
\end{align}
The simple pole $z_I$ for this diagram is given by 
\begin{align}
z_I=\frac{m\langle 23\rangle}{\langle 3|p_1|2]}~.
\end{align} 
After a bit of algebra the $\widehat{I}$-dependent terms can be removed and then one can write the Compton amplitude as 
\begin{align}
\mathcal{A}_{4}\left[\textbf{1},2^+,3^-,\textbf{4}\right] =\frac{g^2}{m^2} \frac{\langle 3|p_4|2]^2[\widehat{\textbf{1}}\textbf{4}]^2}{s_{23}(\widehat{s}_{24}-m^2)}~.
\end{align}
To complete the computation we need to compute the ``shifted" spinor product at $z\, =\, z_{I}$.
 Consider $(\widehat{s}_{24}-m^2)=\langle \widehat{2}4^J\rangle [4_J2]$. The shifted spinor product can be computed as follows
\begin{align*}
\langle \widehat{2}4^J\rangle &=\langle 24^J\rangle +\frac{\langle 32\rangle \langle 4^J|p_1|2]}{\langle 3|p_1|2]} \\
&=\frac{[1_I2]}{\langle 3|p_1|2]} \left( \langle 24^J\rangle \langle 31^I\rangle +\langle 32\rangle \langle 4^J1^I\rangle \right)~.
\end{align*}
Using the Schouten identity 
\begin{align}
\langle 24^J\rangle \langle 31^I\rangle +\langle 23\rangle \langle 1^I4^J\rangle +\langle 1_I2\rangle \langle 34^J\rangle=0~,
\end{align}
and we can write
\begin{align}
\langle \widehat{2}4^J\rangle &=\frac{(s_{12}-m^2)\langle 34^J\rangle}{\langle 3|p_1|2]} ~\Rightarrow  (\widehat{s}_{24}-m^2)=-(s_{12}-m^2)~.\label{interfour1}
\end{align}
Similarly we have
\begin{align}
[\widehat{1}^I4^J]&=\frac{m}{\langle 3|p_1|2]} \left(\langle 31^I\rangle [24^J] +[21^I]\langle 34^J\rangle \right)~.
\end{align}
In terms of bold notation, we have
\begin{align}
[\widehat{\textbf{1}}\textbf{4}]^2=\frac{m^2}{\langle 3|p_1|2]^2}\left( \langle 3\textbf{1}\rangle [2\textbf{4}]+[2\textbf{1}]\langle 3\textbf{4}\rangle \right)^2 ~.\label{interfour2}
\end{align} 
Using equations \eqref{interfour1} and \eqref{interfour2}, we obtain the final expression for the four-particle amplitude: 
\begin{align}
\mathcal{A}_{4}\left[\textbf{1},2^+,3^-,\textbf{4}\right] =g^2 \frac{\left( \langle 3\textbf{1}\rangle [2\textbf{4}]+[2\textbf{1}]\langle 3\textbf{4}\rangle \right)^2}{s_{23}(s_{12}-m^2)} ~.\label{fourparticlespinone}
\end{align}
The result precisely matches with the result obtained directly by unitarity methods \citep{Arkani-Hamed:2017jhn}.
We can similarly compute the Compton amplitude for different helicity configurations of gluons. For example it can be readily checked that
\begin{align}
\mathcal{A}_4\left[\textbf{1},2^+,3^+,\textbf{4}\right]=g^2\frac{[23]^2 \langle \textbf{1}\textbf{4}\rangle ^2}{s_{32}(s_{12}-m^2)}~, \label{fourparticlespinone2}
\end{align} 
which can agrees with the results of \citep{Badger:2005zh}.

\subsection{Massive spin-$1$ with gluon: Five-particle amplitude}\label{section: 4.4}

So far we have computed results that have been previously computed in the literature but using the new recursion relations and this gives us confidence in the validity of our methods. We next compute the colour-ordered five-point amplitude involving two massive spin-$1$ particles and three gluons with arbitrary helicity using the $[\textbf{2}3 \rangle$ shift. This is will lead to the first new result using these methods. The recursion relation for computing this amplitude is given by:
\begin{align}
\mathcal{A}_5[\textbf{1}^1,\textbf{2}^1,3^{h_1},4^{h_2},5^{h_3}]=\sum _{I} \widehat{\mathcal{A}}_{3}(z_I) \frac{1}{P^2-m^2}\widehat{\mathcal{A}}_{4} (z_I)+\sum _{J} \widehat{\mathcal{\tilde{A}}}_{3}(z_J) \frac{1}{P^2}\widehat{\mathcal{\tilde{A}}}_{4} (z_J)~.\label{higgsfivepoint}
\end{align}
Although the final expression (even for such a lower-point amplitude) is rather complicated, we will verify that our result will go over to that of the known massless result in the high energy limit. We have to compute the following diagrams:
\begin{figure}[H]
\includegraphics[scale=.28]{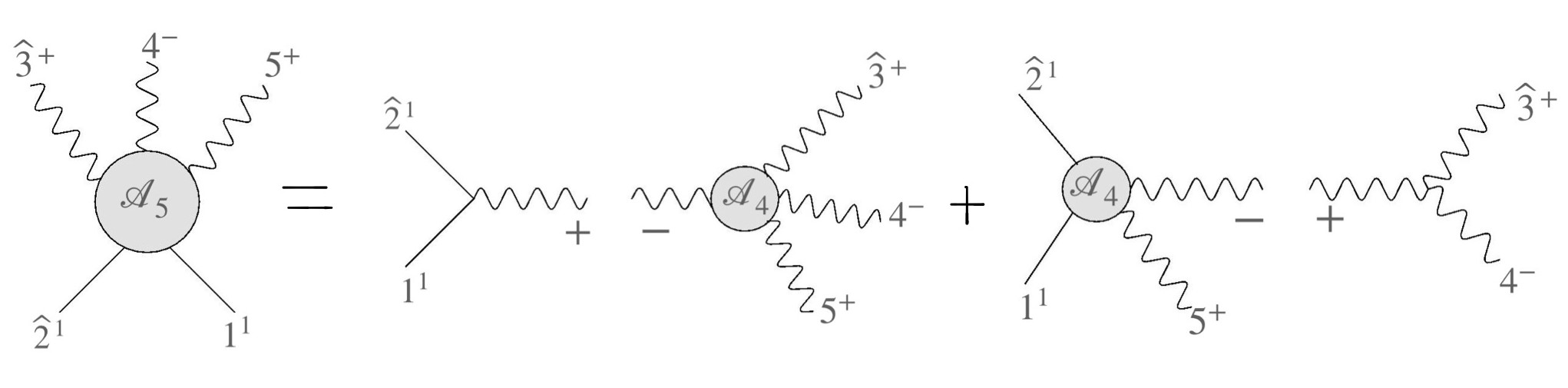}
\caption{Five-particle amplitude in Higgsed Yang-Mills}
\label{figure5}
\end{figure}
\noindent
As in scalar QCD, there exist two non-vanishing scattering channels with massless propagator for this process. The diagrams with a massive propagator will again vanish due to the fact that a massive spin-$1$ particle can not decay into two helicity-$1$ particles. We therefore have to consider only those diagrams in figure \eqref{figure5}.

The contribution from the first diagram can be written using the recursion scheme \eqref{higgsfivepoint} as follows:
\begin{align}
\mathcal{A}_5^{(I)}\left[\textbf{1},\textbf{2},3^+,4^-,5^+\right]=\widehat{\mathcal{A}}_3\left[ \textbf{1},\widehat{\textbf{2}},\widehat{I}^+\right]\frac{1}{s_{12}}\widehat{\mathcal{A}}_4\left[\widehat{I}^-,\widehat{3}^+,4^-,5^+\right].
\end{align}
The three- and four-particle sub-amplitudes have to be evaluated at $z=z_I$ which is
\begin{align}
z_I=\frac{m^3+mp_1\cdot p_2}{\langle 1^I|p_2|3][31_I]}~,
\end{align} 
for the first diagram. The three-point sub-amplitude can be written by using standard methods described in \citep{Arkani-Hamed:2017jhn}
\begin{align}
\widehat{\mathcal{A}}_3\left[ \textbf{1},\widehat{\textbf{2}},\widehat{I}^+\right]=\frac{g}{m}\widehat{x}_{12}\langle \textbf{1}\textbf{2}\rangle ^2=g\frac{[\widehat{I}3]}{\langle \widehat{I}|p_1|3]}\langle \textbf{1}\textbf{2}\rangle ^2.
\end{align}
The four-particle amplitude is a massless amplitude for gluons which is given by the Parke-Taylor formula: 
\begin{align}
\widehat{\mathcal{A}}_4\left[\widehat{I}^-,\widehat{3}^+,4^-,5^+\right]=\frac{[35]^4}{[\widehat{I}3][34][45][5\widehat{I}]}~.
\end{align}
After solving for the intermediate spinor-helicity variable $\widehat{I}$ and evaluating the shifted spinor products at $z_I$, the contribution of the first diagram to the five-particle amplitude is
\begin{align}
\mathcal{A}_5^{(I)}\left[\textbf{1},\textbf{2},3^+,4^-,5^+\right]=g^3\frac{\langle \textbf{1}\textbf{2}\rangle ^2 [53]^4 }{([3|p_2\cdot p_1|5]+m^2[35])[45][34]s_{12}}~. \label{fivepointfirst}
\end{align}

The second diagram that contributes to the five-particle amplitude can be again written using three- and four-particle sub-amplitudes as follows:
\begin{align}
\mathcal{A}_5^{(II)}\left[\textbf{1},\textbf{2},3^+,4^-,5^+\right]=\widehat{\mathcal{A}}_4\left[\textbf{1},\textbf{2},\widehat{I}^-,5^+\right] \frac{1}{s_{34}}\widehat{\mathcal{A}}_3\left[\widehat{I}^+,\widehat{3}^+,4^-\right]~.
\end{align}
The simple pole $\tilde{z}_I$ for second diagram is given by 
\begin{align}
\tilde{z}_I=\frac{m\langle 34\rangle}{\langle 4|p_2|3]}~.
\end{align}
We will not give all the steps for computing the subamplitudes in terms of undeformed spinor-helicity variables as they are more or less similar to the previous calculation. The contribution from second diagram to five-particle amplitude is
\begin{align}
\mathcal{A}_5^{(II)}\left[\textbf{1},\textbf{2},3^+,4^-,5^+\right]=-g^3 \frac{ \left[ \langle 4|p_2|3][5\textbf{1}]\langle 4\textbf{2}\rangle +\langle \textbf{1}4\rangle \left\lbrace\langle 4|p_2|3][5\textbf{2}]+\langle 4|p_3|5][3\textbf{2}]\right\rbrace \right]^2}{\langle 43\rangle \langle 45\rangle (s_{32}-m^2)(s_{15}-m^2) \big([3|p_2\cdot p_1|5]+m^2[35] \big)} \label{fivepointsecond}
\end{align}
The full colour-ordered five-particle amplitude is obtained by adding \eqref{fivepointfirst} and \eqref{fivepointsecond}.

\subsection{High energy limit}

The finite energy amplitude contains contributions from all possible helicity configurations of the massive spin-$1$ particle. This can be seen by expanding any spinor product involving massive spinor-helicity variable using \eqref{hevar1} and \eqref{hevar2}. For example if we have the following expansion
\begin{align}
\langle 3\textbf{1} \rangle ^2 =\langle 3\lambda _1\rangle ^2 \xi ^{-I_1}\xi ^{-I_2}+\langle 3\eta _1\rangle ^2 \xi ^{+I_1}\xi ^{+I_2}+\langle 3\lambda _1\rangle \langle 3\eta _1\rangle \left( \xi ^{-I_1}\xi ^{+I_2}+\xi ^{+I_1}\xi ^{-I_2} \right)~,
\end{align}
then the high energy contribution for different helicity configurations  are given by
\begin{align}
\langle 3\lambda _1\rangle ^2~:\quad &\text{$(-)$ helicity}\\
\langle 3\eta _1\rangle ^2~:\quad &\text{(+) helicity} \\
\langle 3\lambda _1\rangle \langle 3\eta _1\rangle~:\quad &\text{longitudinal}~.
\end{align}
As $\eta _1$ vanishes in the high energy limit, the spinor product $\langle 3\textbf{1}\rangle ^2$ survives only when the massive particle has $(-)$ helicity in high energy limit. Similarly the spinor product $[3\textbf{1}]^2$ survives when the massive particle attain $(+)$ helicity in the high energy limit. 

To check the consistency of the finite energy result for the five-point amplitude, we  take its high energy limit and compare it with the known Parke-Taylor formula for gluon amplitudes. Consider the helicity configuration $(1^-,2^-)$ for the massive particles. Using the procedure laid out earlier this section, we immediately conclude that only the first diagram will contribute to this massless amplitude:
\begin{align}
\mathcal{A}_5^{(HE)}\left[1^-,2^-,3^+,4^-,5^+\right]=g^3\frac{[35]^4 }{[12] [23][34][45][51]}~.
\end{align}
From the structure of finite energy amplitude it is evident that the $(1^+,2^+)$ helicity configuration has vanishing massless amplitude. Next we consider the $(1^+,2^-)$ configuration for which only the second diagram contributes
\begin{align}
\mathcal{A}_5^{(HE)}\left[1^+,2^-,3^+,4^-,5^+\right]=g^3\frac{\langle 24\rangle ^4}{\langle 12 \rangle \langle 23\rangle \langle 34\rangle \langle 45\rangle \langle 51\rangle}~.
\end{align}
Similarly the second diagram produces the correct massless amplitude with $(1^-,2^+)$ helicity configuration. We thus find the correct expected behaviour of the amplitudes in the high energy limit. 

For the case in which all gluons have positive helicity, we consider the same massive-massless shift $[\textbf{2}3\rangle $ and obtain the following expression for five-particle amplitude
\begin{align}
\mathcal{A}_5[\textbf{1},\textbf{2},3^+,4^+,5^+]=g^3\frac{\langle \textbf{1}\textbf{2}\rangle ^2 (\langle 4|p_2|3]s_{12}+\langle 34\rangle [3|p_1\cdot p_2|3])}{\langle 54\rangle ^2 \langle 34\rangle (s_{15}-m^2)(s_{23}-m^2)}~.
\end{align}

\section{Validity of Massive-Massless Shift}\label{section:5}

The examples presented here clearly suggest that the proposed massive-massless shift is a valid shift for computing amplitudes involving massive particles. In this section we present a proof for the validity of such a massive-massless shift by studying the behaviour of the deformed amplitudes for large $z$ in  i) the Higgsed Yang-Mills theory and ii) massive scalar QCD. Since the proof is technical we begin with a brief summary of the main ideas involved in the proof.


Our proof is inspired by the analogous proof in \cite{Nima2008}. However conceptually there is a key difference that we highlight below. The proof regarding the validity of the BCFW shift for massless particles considered a set up where a highly boosted gluon was scattered off a background of low energy massless fields. For real momenta, this corresponds to the familiar Eikonal scattering. The background was referred to as a soft background. In the Eikonal approximation, the helicity of the highly boosted particle was conserved. This conservation law was shown to be a consequence the so-called spin-Lorentz symmetry which was then used to constrain the amplitude at large $z$.

In our case, the soft background is in fact replaced by a static background which includes a collection of massive  and soft massless particles. Our set up is hence closer to the scattering of a boosted gluon off a heavy scattering center surrounded by a cloud of soft gluons. As we show, the resulting outgoing states are a highly boosted massive spin-1 boson and a highly boosted gluon. At infinite boost, the dominant contribution to the amplitude at large $z$ is when the helicity of the boosted gluon is unchanged. Thus, as in the case of massless theories, this contribution is constrained by the spin-Lorentz symmetry. We then use the Ward identity for massless gluons to constrain the sub-leading behaviour of the amplitude and show that for a particular class of shifts, which we refer to as valid shifts, the amplitude vanishes as $\frac{1}{z}$ for large $z$.

\subsection{Higgsed Yang-Mills Theory}

In order to check the validity of massive-massless shift of the type $[\textbf{m}+\rangle$ for the generalized recursion \eqref{BCFWrecursion}, we consider two-point natural amplitude $\widehat{\mathcal{A}}^h_{IJ}$ involving one gluon with helicity $h$ and a massive spin-1 particle with little group indices $(I,J)$. This can be interpreted as a highly boosted gluon scattered through a static background, producing a boosted massive spin-1 particle in the out state or vice-versa.  The validity of the recursion requires
\begin{align}
\widehat{\mathcal{A}}^h_{IJ} =0 ~ \quad \text{for $z \rightarrow \infty$}~. \label{vanishingoftwo}
\end{align}
The three-point amplitude (which is the basic building block of amplitudes using the recursion method) can be constructed from this two-point amplitude by attaching an unshifted external momentum. Since this will also vanish at large $z$ once the two-point amplitude vanishes, any $n-$point amplitude also vanishes once the condition \eqref{vanishingoftwo} is satisfied.

The two-point natural amplitude $\widehat{\mathcal{A}}^h_{IJ}$ is obtained from the field theoretic (or Feynman) amplitude $\widehat{\mathcal{A}}^{\mu\nu}$ (\citep{Arkani-Hamed:2017jhn}) by contracting the latter with the polarization tensors of the gluon and massive vector boson:

\begin{align}
\widehat{\mathcal{A}}^h_{IJ}=\widehat{\mathcal{A}}^{\mu\nu}\,\widehat{e}^h_\mu(j)\,\widehat{e}_{\nu IJ}(i).
\label{Anatural}
\end{align}
At large $z$ the two-point natural amplitude in \eqref{Anatural} can be split into three parts depending on the three modes of polarization of the massive vector boson. We shall call these modes the transverse$(\pm)$ and longitudinal modes. 

\subsubsection{Structure of $\widehat{\mathcal{A}}^{\mu\nu}$ \label{section 5.1.1}}
In this section we find the tensor structure of the field theoretic amplitude  $\widehat{\mathcal{A}}_{\mu\nu}$ for the case of
a scalar field coupled to a Yang-Mills field and an abelian gauge field. This theory produces the three-point interaction in \eqref{minimalthreepoint} between the photon and the massive vector boson after Higgsing \cite{johansson19}. The Lagrangian for this theory is
\begin{align}
	\mathcal{L}=-\frac{1}{2}\Tr\left(F_{\mu \nu}F^{\mu \nu} \right)-\frac{1}{4}\left(B_{\mu \nu}B^{\mu \nu} \right) +\frac{1}{2} \left( D_\mu \Phi \right)^\dagger D^\mu \Phi   ~.\label{Weakint}
\end{align}
Here the field strengths $F_{\mu \nu}$ and $B_{\mu \nu}$  are associated with $SU(2)$ gauge field $A_\mu ^a$ and $U(1)$ gauge field $B_\mu$   respectively. Next we do a background field expansion of the gauge fields before spontaneous symmetry breaking as this requires manifestly gauge invariant Lagrangian. We will see later that this procedure indeed gives the correct interaction term that is suitable for a constructible generalized recursion relation. Expanding the gauge fields as  sum of background and fluctuation fields:
\begin{align}
A^c_\mu =A^c_{0\mu} +a^c_\mu ~;\qquad B_\mu =B_{0\mu }+b_\mu~,
\end{align}
the field strength for the non-abelian gauge field takes the following form:
\begin{align}
F_{\mu \nu}^c&=F_{\mu \nu}^c(A_0)+D_{A[\mu}a_{\nu ]}^c-ig\epsilon ^{cde}a_{d\mu}a_{e\nu}~.
\end{align}
Here we have introduced the background field $A_0$-covariant derivative as follows
\begin{align}
D_{A\mu}a^c_\nu= \partial _\mu a^c_\nu -ig\epsilon^{cde}A_{0d\mu}a_{e\nu}
\end{align}
Similarly the field strength for the abelian gauge field can be written as:
\begin{align}
B_{\mu \nu}=B_{\mu \nu}(B_0)+B_{\mu \nu}(b).
\end{align}
The gauge covariant derivative appearing in the scalar kinetic term can be expanded as 
\begin{align}
	D_\mu \Phi  =\left(\partial _\mu \Phi  -ig(A^m_{0\mu}+a^m_\mu) \tau_m\Phi  -\frac{ig^{'}}{2} (B_{0\mu}+b_\mu) \Phi  \right) ~.
\end{align}
We will be interested in terms which are quadratic in $a_\mu$ as it turns out that only they generate terms involving the product of massive and massless fields after spontaneous symmetry breaking.
These terms arise from the kinetic term of the $SU(2)$ gauge field:
\begin{align}
-\frac{1}{4}\left( F_{\mu \nu}^c\right)^2 \rightarrow -\frac{1}{2}\left( D_{A\mu}a_{\nu}^c D_{A}^{\mu}a^{\nu }_c \right)+\frac{i}{2}g\epsilon ^{cde}a_{d\mu}a_{e\nu}F_{c}^{\mu \nu}(A_0)~, \label{releventterms1}
\end{align}
where we have used the gauge fixing condition $D_{A\mu}a^\mu=0$. Note that we could also consider the kinetic term for the abelian gauge field $(b_\mu)$. But in that case, we will see later on that they do not lead to terms involving the  massive gauge fields after Higgsing.

Here the fluctuation fields $a^c_\mu$ and $b_\mu$ are massless. In order to get a massive particle in the external state of the two-point amplitude we use the Higgs mechanism as this is the only way to generate mass for non abelian gauge fields. Since gauge invariance of the Lagrangian remains intact after Higgsing (though it is not manifest), this allows us to use the Ward identity which turns out to be crucial in determining the large $z$ behaviour of the amplitude.

After Higgsing, the massless fluctuation  fields $\left\lbrace a^c_\mu \right\rbrace$ and $b_\mu$ can be written in terms of newly defined massive fields $(w^+_\mu,w^-_\mu)$ and massless field $(u_\mu)$ as follows
\begin{align}
a^1_\mu =\frac{1}{\sqrt{2}}(w^+_\mu +w^-_\mu)~,\quad a^2_\mu =\frac{i}{\sqrt{2}}(w^+_\mu -w^-_\mu)~, \quad a^3_\mu =\frac{\sqrt{g^2+g^{'2}}}{g^{'}} \left( u_\mu -\frac{g}{\sqrt{g^2+g^{'2}}} b_\mu\right)~. \label{FieldRedefinition}
\end{align}

Recall that we are looking at a process where a highly boosted gluon is scattered off a static background of massive spin-1 particles and soft gluons. So to obtain the field theoretic amplitude $\widehat{\mathcal{A}}^{\mu\nu}$ first we look for terms containing $(w^-_\mu u_\nu)$ in the Lagrangian which accounts  for the interaction of the massive vector boson and the photon.  From the kinetic term \eqref{releventterms1} and using the definitions of massive and massless Higgsed fields following \eqref{FieldRedefinition}, we can write the relevant terms as
\begin{align}
\mathcal{L}^{w^-;u}
&= \frac{i}{2 \sqrt{2}\tilde{g}} (F_{2}^{\mu \nu}(A_0)-F_{1}^{\mu \nu}(A_0)) w^-_\mu u_\nu~,
\end{align}
where $1,2$ refer to the colour degrees of freedom, $F_{1,2}^{\mu \nu}(A_0)$ are the field strengths for background gauge field $A_0^\mu$ and 
\begin{align}
\tilde{g}=\frac{g^{'}}{g\sqrt{g^2+g^{'2}}}~.
\end{align} 
Note that we have two massive fields $w^{\pm}_\mu$ after symmetry breaking, representing two different massive particles. For our present purposes, we need only one of them. Of course, the final conclusion of this section does not depend on this choice. 
Up to now we have considered only interactions involving the massless abelian gauge field $(u_\mu)$. To account for the three-point interaction involving the gluon and the massive spin-1 particle, the above Lagrangian can be uplifted to its non-abelian version as follows:
\begin{align}
\mathcal{L}^{w^-;u}
&= \frac{i}{2 \sqrt{2}\tilde{g}} \Tr \left[(F_{2}^{\mu \nu}(A_0)-F_{1}^{\mu \nu}(A_0)) w^-_\mu u_\nu\right]~.
\end{align}

So far we have studied those terms in the Lagrangian that are relevant for proving the validity of the massive-massless shift. However we can also easily include the terms that are required to prove the validity of  massless-massless shift by keeping track of the terms quadratic in the massless Higgsed field $u_\mu$. There are potentially three sources for these terms: i) the kinetic term for the $SU(2)$ gauge field: $D_{A\mu}a_{3\nu}D_A^\mu a_3^\nu $, ii) the kinetic term for the $b^\mu$ field \footnote{Note that after symmetry breaking, we should treat this term as non-abelian field strength of field $u_\mu$.}: $B_{\mu \nu }(b)B^{\mu \nu}(b)$ and iii) the kinetic term for the scalar field $\Phi$. Taking into account all these terms, the gauge fixed Lagrangian relevant for both types of shifts is given by:
\begin{align}
\mathcal{L}&=\frac{i}{2 \sqrt{2}\tilde{g}} \Tr \left[(F_{2}^{\mu \nu}(A_0)-F_{1}^{\mu \nu}(A_0)) w^-_\mu u_\nu\right]\cr
&+\Tr\left[ -\frac{g^2+g^{'2}}{2g^2}D_{A\mu}u_\nu D_{A}^\mu u^\nu -\frac{g^2+g^{'2}}{4g^2}\partial _\mu u_\nu \partial ^\mu u^\nu +\frac{g^{'2}}{4g^2}(g^2+g^{'2})\phi _0^2 u_\mu u^\mu \right], \label{BCFWLagrangian}
\end{align}
where we fix the gauge freedom of $u^\mu$ by setting $\partial _\mu u^\mu =0$ and $\phi _0\,$is the v.e.v of the scalar field. We would like to point out that for the case of massive-massive shifts, one would consider terms that are quadratic in the massive Higgsed fields $(w^{\pm}_\mu)$. Since such shifts are outside the scope of this work, we have omitted such terms in  \eqref{BCFWLagrangian}. 

Due to the introduction of the non-vanishing background  fields the usual Lorentz symmetry of this Lagrangian is explicitly broken. However following the discussion in \citep{Nima2008} we note that the terms in the second parenthesis are such  that the indices of the fluctuation fields are contracted with each other and these terms have an enhanced symmetry (the so called spin Lorentz symmetry) - a Lorentz transformation that acts on the $\nu$ indices of the fluctuation fields $u_\nu$. We will use this symmetry to constrain the form of the two-point amplitude $\widehat{\mathcal{A}}^{\mu\nu}$ at large $z$.  To make this symmetry explicit, we simply re-label the indices of the fluctuation fields and write the Lagrangian as
\begin{align}
\mathcal{L}
=&\frac{i}{2 \sqrt{2}\tilde{g}}\Tr \left[(F_{2}^{c d}(A_0)-F_{1}^{c d}(A_0)) w^-_c u_d\right] \nonumber \\
&+\Tr \left[ -\frac{g^2+g^{'2}}{2g^2}D_{A\mu}u_c D_{A}^\mu u^c -\frac{g^2+g^{'2}}{4g^2}\partial _\mu u_c \partial ^\mu u^c +\frac{g^{'2}}{4g^2}(g^2+g^{'2})\phi _0^2 u_c u^c\right] ~.
\label{interactionLag}
\end{align}

The  contribution to the amplitude due to the derivative terms in the second parenthesis of the Lagrangian is dominant ($\mathcal{O}(z)$) at the large $z$ limit  and is proportional to $\eta _{cd}$ due to it's spin Lorentz invariance. The repeated use of these vertices will contribute higher powers in $z$ to the amplitude. The third term in the second parenthesis gives sub-leading contribution to the amplitude at the large-$z$ limit and is also proportional to $\eta _{cd}$. The two terms in the first parenthesis  that have the background fields explicitly break this symmetry and so the contribution of the single insertion of these vertices is  proportional either to the field strengths $F^{cd}_{1,2}(A_0)$ or a combination of these, so that the contribution is anti symmetric in spin-Lorentz indices. Further insertions of these vertices contribute additional powers in $\frac{1}{z}$ multiplying general matrices. Thus by utilizing the spin Lorentz symmetry that dominates the large-$z$ behaviour leads us to infer the following matrix structure for the two-point amplitude:
\begin{align}
\widehat{\mathcal{A}}^{c d}_{\text{Full}}=\eta ^{cd}(a+bz+ \ldots )+M^{c d} +\frac{1}{z}\left(\tilde{B}^{cd}+B^{c d}\right)+\mathcal{O}\left(\frac{1}{z^2}\right)~, \label{ABCFW}
\end{align}
where $M^{c d}$ is an anti-symmetric matrix written in terms of the background field strengths, and $B^{c d}$ and $\tilde{B}^{cd}$ are general matrices. 

Now we separately discuss the validity of the two types of generalized shifts. The structure of two-point amplitude with boosted massive and massless particles can be extracted from the above amplitude
\begin{align}
\widehat{\mathcal{A}}^{c d}_{\text{massive-massless}}=M^{c d} +\frac{1}{z}B^{c d}+ \ldots ~~~,\label{vaertices}
\end{align}
The  two-point amplitude with only boosted massless particles as external particles is
\begin{align}
\widehat{\mathcal{A}}^{cd}_{\text{massless-massless}}=\eta ^{cd}(a+bz+ \ldots)+\frac{1}{z}\tilde{B}^{cd}+ \ldots
\end{align}
These expressions will prove to be crucial in proving the validity of the generalized shifts.

We postpone discussion of the massless-massless shifts to Section \ref{section:6}. In the following subsections, we first concentrate on the validity of massive-massless shift and work with the two-point amplitude in \eqref{vaertices}. 
To check the validity of the shift of type $[\textbf{m}+\rangle $, we consider the case where gluon has positive helicity.
 Using \eqref{map} and \eqref{Anatural} we can write the natural amplitude as
\begin{align}
\widehat{\mathcal{A}}^+_{IJ}=
\begin{cases}
\widehat{\mathcal{A}}^+_{\pm} = \kappa \widehat{\mathcal{A}}^{ab}r_a\widehat{e}_{b\pm}(i)  & \text{for transverse modes,}\\
\widehat{\mathcal{A}}^+_{0}=\kappa \widehat{\mathcal{A}}^{ab}r_a\widehat{e}_{b0}(i)  & \text{for longitudinal modes.}\\
 \end{cases}
 \label{onshell}
\end{align}

However the large-$z$ limit of the amplitude combined with the action of the Ward identity acts differently on the  longitudinal and transverse modes. 
For the longitudinal mode, we use a result proved in \citep{Cohen:2010mi}, that the large-$z$ limit together with the Ward identity for spontaneously broken gauge theory allows us to treat the amplitude for this mode as massless scalar-gluon amplitude. Whereas, the amplitude for  the transverse modes can be treated as an amplitude involving only gluons in the large-$z$ limit. 

\subsubsection{Validity for transverse modes}

Let us consider the transverse modes of the amplitude in the $z \rightarrow \infty $ limit. In this limit, as discussed previously, the deformed particles are highly boosted and can be considered as massless. The on-shell deformed amplitude $\widehat{\mathcal{A}}^{ab}$ satisfies the Ward identity for Yang-Mills theory:
\begin{align}
	\widehat{p}_{ja} \widehat{\mathcal{A}}^{ab}\widehat{e}^{\pm}_{b}(i)=0~. \label{wardid}
\end{align}
Here we denote $\widehat{\mathcal{A}}^{ab}_{\text{Massive-massless}}$ simply as $\widehat{\mathcal{A}}^{ab} $ to avoid clutter. Using the Ward identity and  
the shift equation for $p_j$ \eqref{shift} we can write 
\begin{align}
r_a \mathcal{A}^{ab}\widehat{e}^{\pm}_b(i)=-\frac{1}{z}p_{ja}\mathcal{A}^{ab}\widehat{e}^{\pm}_b(i).
\label{Award}
\end{align}
 Using \eqref{vaertices}, \eqref{onshell} and \eqref{Award},  the transverse modes of the on-shell amplitude has the following behaviour in the large-$z$ limit:
\begin{align}
\left. \widehat{\mathcal{A}}^+_{\pm} \right|_{z\rightarrow \infty}=-\frac{\kappa}{z}\left[M^{ab} +\frac{1}{z}B^{ab}+ \ldots \right] p_{ja}\widehat{e}_{b}^{\pm}(i).
\end{align}
To evaluate these we need to analyze the large-$z$ behaviour of the deformed polarization vectors $\widehat{e}_{b}^{\pm}(i)$. By choosing the reference spinor $\zeta ^\alpha =\lambda ^\alpha _j$, the positive helicity polarization vector can be written as follows:
\begin{align}
\widehat{e}^+_b(i)= \Sigma_{ijb}-\frac{z}{m}p_{jb}\,\Omega_{ij}~,\label{plusmap}
\end{align}
where we have defined 
\begin{align}
\Sigma _{ijb}=\frac{\langle \lambda _j|\sigma _b|\tilde{\lambda}_i]}{\sqrt{2}\langle \lambda _j \lambda _i\rangle} \quad\text{and}\quad \Omega_{ij}=\frac{[\tilde{\lambda}_i \tilde{\lambda}_j]}{\sqrt{2}\langle \lambda _j\lambda _i \rangle}~.
\end{align}
We refer the reader to appendix \ref{appendix:b} for more details. Similarly the negative helicity component of massive polarization can be written in the following way by choosing the reference spinor $\zeta _{\dot{\alpha}}=\tilde{\lambda}_{j\dot{\alpha}}$:
\begin{align}
\widehat{e}^-_b(i)|_{z \rightarrow \infty}= \frac{\Sigma _{ijb}^*}{\sqrt{2}[\tilde{\lambda}_i\tilde{\lambda}_j]}=\Upsilon_{ijb}~.\label{minusmap}
\end{align}
Using \eqref{plusmap} and \eqref{minusmap} the on-shell amplitudes have the following limiting behaviour as $z \rightarrow \infty $:
\begin{align}
\widehat{\mathcal{A}}^+_{-} =0 ~,\qquad \widehat{\mathcal{A}}^+_{+} =\frac{k}{m}\Omega_{ij}\,M^{ab}\,p_{ja}p_{jb}=0 \label{transverse} ~~,
\end{align}
where we have used $M^{ab}=-M^{ba}$.

We thus find that both the natural amplitudes $\widehat{\mathcal{A}}^+_{\pm}$ vanish in the large-$z$ limit, thereby proving the validity of the massive-massless shift  for the transverse modes.

\subsubsection{Validity for longitudinal mode}\label{section:longt}
In this section we will analyse the large-$z$ behaviour of $\widehat{\mathcal{A}}_0^+$ involving the longitudinal mode of vector boson. The amplitude $\widehat{\mathcal{A}}_0^+$  is related to the amplitude involving massless scalars and gluons via the Ward identity for the spontaneously broken gauge theory \citep{Cohen:2010mi}. So to find the large-$z$ behaviour of $\widehat{\mathcal{A}}_0^+$ it is sufficient to analyse the large-$z$ behaviour of the  amplitude involving massless scalars and gluons.

 Let us consider the following four-particle amplitude involving two massless scalars and gluons:
\begin{center}
	\includegraphics[scale=.3]{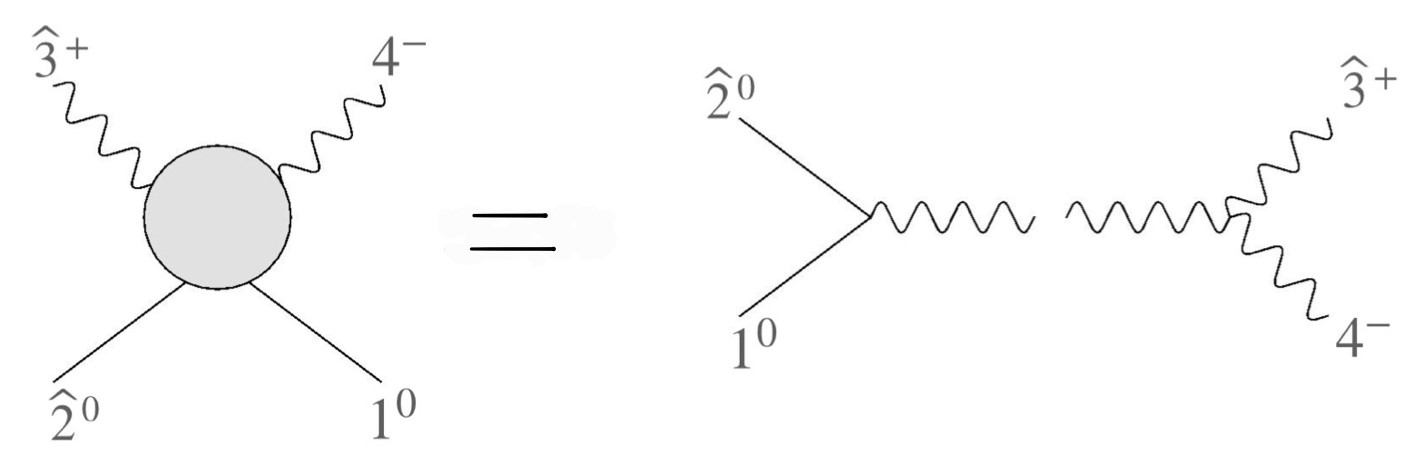}
\end{center}
For the vector boson amplitude, we considered the shift in the adjacent legs , so we have to  shift the adjacent scalar-gluon legs in this case.
The scalar is in the adjoint representation of the gauge group and we consider colour-ordered amplitudes. The relevant three-point interactions are encoded in the following interaction Lagrangian:
\begin{align}
\mathcal{L}_{3}=ig  \left[ (\partial _\mu \phi) A^\mu \phi ^*-A_\mu \phi \partial ^\mu \phi ^*-\frac{1}{2}\Tr (\partial ^\mu A^\nu)[A_\mu, A_\nu] \right]. \label{lagrangian}
\end{align}
 We expand both the scalar and vector fields in terms of the background and fluctuation fields as:
\begin{align}
	\phi =\phi _0 +\xi ~;\qquad A^\nu =A_0^\nu +a^\nu ~;\quad \phi ^*=\phi ^*_0 +\xi ^*~.
\end{align}
Next we collect the $\mathcal{O}(z)$ terms from the Lagrangian and show that they can be made $\mathcal{O}(1)$  by using the gauge condition $\partial_\mu a^\mu=0$. The  $\mathcal{O}(z)$ terms in the Lagrangian \eqref{lagrangian} are given by
\begin{align}
	\mathcal{L}_{3}&\supset ig \big( (\partial _\mu \xi)\,a^\mu \phi ^*_0-a_\mu \phi _0\partial ^\mu \xi ^* -\Tr(\partial ^\mu a^\nu)[a_\mu, A_{0\nu} ] \big)~.
\end{align} 
The first two terms can be made $\mathcal{O}(1)$ by integrating by parts and then using the gauge fixing condition ($\partial_\mu a^\mu=0$):
\begin{align}
\partial _\mu \xi a^\mu \phi ^*_0 \sim -\xi a^\mu \partial _\mu \phi ^*_0~\quad , \quad a_\mu \phi _0\partial ^\mu \xi ^*\sim - a_\mu \xi ^*\partial ^\mu \phi _0~.
\end{align}
Similarly the third term can be made $\mathcal{O}(1)$\footnote{Note that when two scalar legs are shifted, the internal propagator is always a scalar and hence the $\mathcal{O}(z)$ vertices contain terms like $(\partial _\mu \xi A^\mu \xi ^* -A_\mu \xi \partial ^\mu \xi ^*)$. These terms cannot made $\mathcal{O}(1)$ by using gauge fixing condition. Hence  shifts involving only scalar legs in Yang-Mills theory do not lead to BCFW type recursion relations.}. So there are no $\mathcal{O}(z)$ vertices when a single scalar and gluon line are shifted. However, the propagator has $\mathcal{O}(1/z)$ behaviour. Hence the overall amplitude has $\mathcal{O}(1/z)$ scaling which vanishes for $z \rightarrow \infty $. All higher-point amplitudes with adjacent scalar-gluon shift will be suppressed by additional $\frac{1}{z}$ factors due to the added propagators. So we conclude that massless scalar-gluon shift is a valid shift to obtain the generalized recursion. Note that  though we have considered colour ordered amplitudes, this colour ordering is not necessary for this proof. From this and the results of \citep{Cohen:2010mi} we infer that the natural amplitude $\widehat{\mathcal{A}}_0^+$  also vanishes at large $z$. 

We have shown that the proposed $[\textbf{m}+\rangle $ shift is a valid shift as the natural amplitude \eqref{onshell} vanishes at large $z$. One can repeat the same  line of arguments to show that the $[-\textbf{m}\rangle $ is also a valid massive-massless shift.

\subsubsection{Comparison with previous results}

In \citep{Franken:2019wqr}, the authors derived conditions to check the validity of  multi line shifts in the context of BCFW type recursion relations. 
The authors studied large-$z$ behaviour of $n$-point scattering amplitudes ($\widehat{\mathcal{A}}(z)\sim z^{\gamma}$) with multi-line shifts and obtained a bound on $\gamma$ that would lead to valid shifts. For the two-line shifts   discussed in this paper, the relevant constraint is given by: 
\begin{align}
\gamma&\leq 1-[\tilde{g}]-s_{\lambda}+s_{\tilde{\lambda}}
\end{align}
where $[\tilde{g}]$ is the mass dimension of coupling, $s_{\lambda}$ and $s_{\tilde{\lambda}}$ are the spin projection ($s_z$) of the particles which are shifted by the spinor-helicity variables ${\lambda}$ and $\tilde{\lambda}$ respectively. For the massive-massless shift we have considered, the ${\lambda}$-variable of the positive helicity gluon is shifted and the $\tilde{\lambda}$-variable of the massive spin-1 particle is shifted. Thus $s_{\lambda}=1$ and $s_{\tilde{\lambda}}=-1$ (min. of $s_z$ for massive spin-1).  

For example consider the four-point amplitude in section \ref{section4.3},
\begin{align}
\widehat{\mathcal{A}}_{4}\left[\widehat{\textbf{1}},\widehat{2}^+,3^-,\textbf{4}\right]=\frac{g^2}{m^2} \frac{\langle 3|p_4|2]^2[\widehat{\textbf{1}}\textbf{4}]^2}{\widehat{s}_{23}(\widehat{s}_{24}-m^2)}~.
\label{zfourpoint}
\end{align}
In this case, we find the condition for valid shifts as $\gamma\leq 1$ where we have used $[\tilde{g}]=-2$. On the other hand, from \eqref{zfourpoint} we get the larg $z$ behaviour of the amplitude as $\mathcal{A}_{4}(z) \sim z^0$. So this satisfies the above validity condition of the shifts. Similarly one can check that for all the amplitudes computed in this paper, the constraints found in \citep{Franken:2019wqr} are satisfied.

\subsection{Scalar QCD}
Now we discuss the proof of validity of the massive-massless shift for scalar QCD. The Lagrangian for the theory of massive scalar QCD \citep{johansson19} is given by
\begin{align}
\mathcal{L}_{\text{scalar~QCD}}= -\frac{1}{4}\Tr\left( F_{\mu \nu}F^{\mu \nu} \right)+\mathcal{L}_{GF}\left(\partial _\mu A^\mu \right)-\frac{1}{2}|D_\mu \phi|^2-\frac{1}{2}m^2|\phi|^2~.
\end{align}
Let us focus on the four-particle amplitude with two massive scalar particles and two gluons. This amplitude can be constructed using three-point scalar gluon vertex and three gluon vertex. The relevant terms are 
\begin{align}
\mathcal{L}_{3}=ig  \left[ (\partial _\mu \phi )A^\mu \phi ^*-A_\mu \phi \partial ^\mu \phi ^*-\frac{1}{2}\Tr(\partial ^\mu A^\nu)[A_\mu, A_\nu] \right]. 
\end{align} 
These terms are exactly same as those we we encountered in section \ref{section:longt} for which we have already proven the validity of the shift. Thus we conclude that massive-massless shift is a valid shift for scalar QCD theory. We mention that if we consider the shift involving only gluons, then also the amplitude vanishes at large $z$, as shown in  \citep{Nima2008}.

\section{Validity of Massless-Massless Shift }\label{section:6}
In this section we prove the validity of  the massless-massless shift for amplitudes involving massive spin-1 particles and gluons. We follow the same line of arguments as in the previous sections. 
Recall that the interaction term responsible for the massless-massless shift is given by the second line in  \eqref{interactionLag}: 
\begin{equation}
{\cal L} \supset \Tr \left[ -\frac{g^2+g^{'2}}{2g^2}D_{A\mu}u_c D_{A}^\mu u^c -\frac{g^2+g^{'2}}{4g^2}\partial _\mu u_c \partial ^\mu u^c +\frac{g^{'2}}{4g^2}(g^2+g^{'2})\phi _0^2 u_c u^c\right]~.
\end{equation}
From this one can infer the structure of two-point amplitude with massless particles as external states. We recall from \eqref{vaertices} that the expression for the two-point amplitude as (in this section alone we denote $\widehat{\mathcal{A}}^{ab}_{\text{massless-massless}}$ as $\widehat{\mathcal{A}}^{ab} $):
\begin{align}
\widehat{\mathcal{A}}^{ab}=\eta ^{ab}(a+bz+ \ldots)+\frac{1}{z}\tilde{B}^{ab}+\ldots
\end{align}

We define the massless-massless shift $[ij\rangle $ in terms of spinor-helicity variables  with the following shift equations
\begin{align}
|\widehat{i}]=|i]-z|j]~;\quad   |\widehat{j} \rangle =|j\rangle +z|i\rangle
\end{align}
The deformed massless polarization can be written in terms of momentum as follows
\begin{align}
\widehat{e}^+_a(i)=\frac{q_a^*-zp_{ja}}{\sqrt{2}\langle ji\rangle} ~,\quad \widehat{e}^+_b(j)=\frac{q_b}{\sqrt{2}\langle ij\rangle}~,\quad \widehat{e}^-_a(i)=\frac{q_a^*}{\sqrt{2}[ij]}~,\quad \widehat{e}^-_b(j)=\frac{q_b^*-zp_{ib}}{\sqrt{2}[ji]}~,
\end{align} 
with $q_{\alpha \dot{\alpha}}=\lambda _{i\alpha}\tilde{\lambda}_{j\dot{\alpha}}$. These relations are exactly same as in the analysis for Yang-Mills theory in \citep{Nima2008}. The massless amplitude in the limit $z\rightarrow \infty $  satisfies the following Ward identity
\begin{align}
\widehat{p}_{ja}\widehat{\mathcal{A}}^{ab}\widehat{e}_b^{\pm}(i)=0 \Rightarrow q_a \widehat{\mathcal{A}}^{ab}\widehat{e}_b^{\pm}(i)=-\frac{1}{z}p_{ja}\widehat{\mathcal{A}}^{ab}\widehat{e}_b^{\pm}(i)\label{Wardid2}
\end{align}
Then the amplitude $\widehat{\mathcal{A}}^+_+$ at large $z$ becomes
\begin{align}
\left. \widehat{\mathcal{A}}^+_+\right |_{z\rightarrow \infty} =-\frac{1}{2\langle ij\rangle ^2} p_{j}^2= 0 ~.
\end{align}
Similar results are obtained for the $[--\rangle$ and $[-+\rangle$ shifts. But for  $[+-\rangle $ shift, the two-particle amplitude does not vanish at large $z$ :
\begin{align}
\widehat{\mathcal{A}}^+_-|_{z\rightarrow \infty}&=\widehat{e}^+_a(i)\widehat{\mathcal{A}}^{ab}\widehat{e}^-_b(i)\cr
&=-\frac{1}{2p_i\cdot p_j}(q_a^*-zp_{ja})\left(\eta ^{ab}(a+bz+..)+\frac{1}{z}\tilde{B}^{ab}+..\right)(q_b^*-zp_{ib})\rightarrow z^3
\end{align}

We conclude that the massless-massless shifts of type $[-+\rangle, [++\rangle, [--\rangle $ are valid, while the $[+-\rangle$ shift is invalid, agreeing with the conclusions of \cite{Nima2008}. However we note that our results are for the Higgsed Yang-Mills theory involving massive particles in the scattering process.

\subsection{Example : Five-particle amplitude}

As a consistency check on our methods and results, in this section, we calculate the same five-particle amplitude as in Section \ref{section: 4.4} but now using the massless-massless shift $[-+\rangle $. 
\begin{center}
\includegraphics[scale=.29]{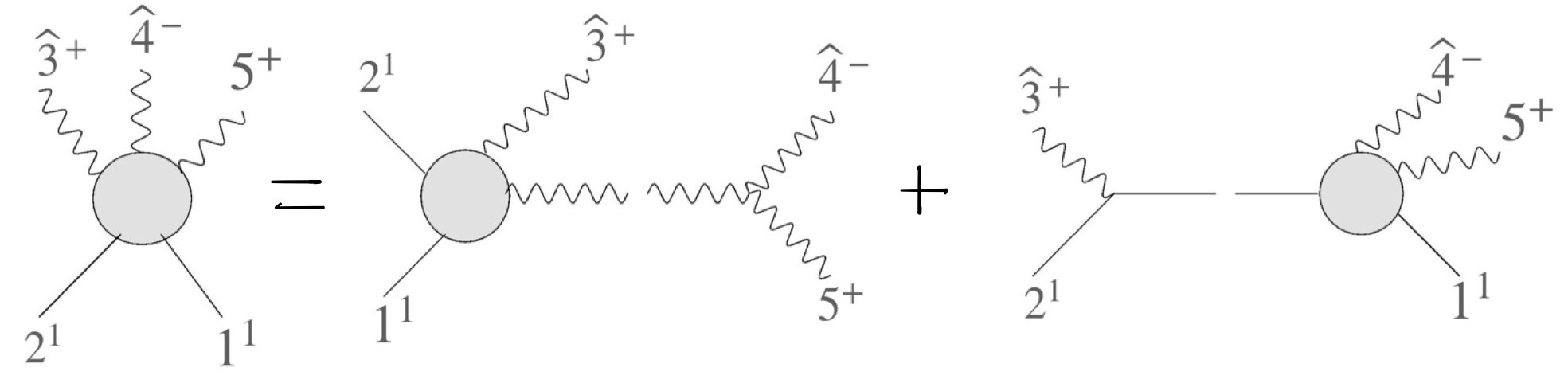}
\end{center}
We shift the $(3^+,4^-)$ gluons and deform the spinor-helicity variables as follows:
\begin{align}
|\widehat{4}]=|4]-z|3]~;\quad |\widehat{4}\rangle =|4\rangle \\
|\widehat{3}\rangle =|3\rangle +z|4\rangle ~;\quad |\widehat{3}]=|3]
\end{align}
Due to this being an adjacent shift there are two possible scattering diagrams. The recursion in \eqref{BCFWrecursion} leads to the following expression for the colour-ordered amplitude: 
\begin{multline}
\mathcal{A}_5[\textbf{1},\textbf{2},{3}^+,{4}^-,5^+]=\mathcal{A}_4[\textbf{1},\textbf{2},\widehat{3}^+,\widehat{I}^+]\frac{1}{{s}_{45}}\mathcal{A}_3[\widehat{I}^-,\widehat{4}^-,5^+] \\
  +\mathcal{A}_4[\widehat{\textbf{I}},\widehat{4}^-,5^+,\textbf{1}]\frac{1}{{s}_{23}-m^2}\mathcal{A}_3[\textbf{2},\widehat{3}^+,\widehat{\textbf{I}}]~.
\end{multline}
The simple poles in the $z$-plane for these two diagrams are located respectively at  
\begin{align}
z_I=\frac{p_4\cdot p_5}{r\cdot p_5}=\frac{[45]}{[35]}\quad\text{and}\quad \tilde{z}_I=-\frac{\langle 3|p_2|3]}{\langle 4|p_2|3]}~.
\end{align}
The full colour-ordered five-point amplitude is given by
\begin{align}
\mathcal{A}_5[\textbf{1},\textbf{2},{3}^+,{4}^-,5^+]&=g^3\frac{\langle \textbf{1}\textbf{2}\rangle ^2 [53]^4 }{([3|p_2\cdot p_1|5]+m^2[35])[45][34]s_{12}}\nonumber \\
&-g^3 \frac{ \left[ \langle 4|p_2|3][5\textbf{1}]\langle 4\textbf{2}\rangle +\langle \textbf{1}4\rangle \left\lbrace\langle 4|p_2|3][5\textbf{2}]+\langle 4|p_3|5][3\textbf{2}]\right\rbrace \right]^2}{\langle 43\rangle \langle 45\rangle (s_{32}-m^2)(s_{15}-m^2) \big([3|p_2\cdot p_1|5]+m^2[35] \big)}~.
\end{align}
The result matches perfectly with the amplitude computed using the massive-massless shift in section \ref{section: 4.4}.

\section{Discussion}\label{section:7}

In this paper we have derived a generalized BCFW recursion relation which includes a complex momentum shift of a massive particle, by making use of the recently proposed spinor-helicity formalism for massive particles \citep{Arkani-Hamed:2017jhn}. We proved the validity of this recursion relation for massive scalar QCD and Higgsed Yang-Mills theories by adapting the background field methods of Arkani-Hamed and Kaplan \citep{Nima2008}. As an explicit check of the new recursion relation we computed the four- and five-particle amplitudes in these theories and found perfect agreement with the results already known in the literature. The five-particle vector boson amplitudes for different helicities of gluons that we obtained with the newly proposed recursion relation is a new result of this formalism. 

The background field expansion allowed us to analyze the validity of all possible massive-massless shifts in scalar QCD and the Higgsed Yang-Mills theories; we show that while the massive-massless shift $[\textbf{m}+\rangle $ is indeed a valid shift, the $[\textbf{m}-\rangle $ shift is invalid. 
For certain particle configurations, it turns out to be more convenient to use massive-massless shift even if massless-massless shifts are in principle allowed as it leads to significant calculational simplification.
Though in principle one can compute all higher-point amplitudes using the BCFW recursion relation for the theories mentioned earlier, it would be quite challenging to find a compact expression for the generic $n$-point amplitude due to intricate structure of lower-point amplitudes itself. Similarly, we have restricted ourselves to the calculation of tree-level amplitudes using the new recursion relation and an obvious question would be to extend these results to loop amplitudes. We leave this challenging technical problem to future work.
 
In the same spirit of massive-massless shift it would be worthwhile to find a shift involving two massive momenta and which is relevant for the computation of all massive amplitudes. Preliminary analysis seems to indicate that the little group covariance of the deformation of massive spinor-helicity variables is difficult to maintain and which makes the computation of amplitudes using massive spinor-helicity formalism less tractable. It would be interesting to explore this in more detail and we defer this to future work.

\vspace{1cm}
\section*{Acknowledgements}

We would like to thank Alok Laddha for suggesting the problem, numerous illuminating discussions and help. We thank him for carefully  going through the draft and providing valuable suggestions and comments on it. We thank Arnab Priya Saha for collaboration at the initial stage of this work and various useful discussions. We thank Sujay Ashok for constant encouragement and valuable comments on the draft. This research was supported in part by the International Centre for Theoretical Sciences (ICTS) for participating in the online program - Recent Developments in S-matrix theory (code: ICTS/rdst2020/07).
\begin{appendix}
\section{Reduction of Massive Polarization}\label{appendix: a}
In this appendix we show how the polarization tensor of massive particles reduces to polarization vectors of massless particles at large $z$. This is an important step in for proving the validity of the massive-massless shift involving the transverse mode of the vector boson. In spinor-helicity formalism, the  polarization tensor for a massive spin-$1$ particle is given by \citep{Guevara:2018wpp, Chung:2018kqs}:
\begin{align}
	e^\mu _{IJ}(i)=\frac{1}{2\sqrt{2}m}\left[ \langle \lambda _{iI}|\sigma ^\mu |\tilde{\lambda}_{iJ}] +(I\leftrightarrow J) \right]~. 
	\label{massivepol1}
\end{align}
The massive spinor-helicity variables can be expanded in terms of pairs of massless spinor-helicity variables $(\lambda _\alpha, \tilde{\lambda}_{\dot\alpha})$, $(\eta _\alpha, \tilde{\eta}_{\dot\alpha})$ as \citep{Arkani-Hamed:2017jhn}:
\begin{align}
	\lambda _{iI}^\alpha &=\lambda _i ^\alpha \xi ^+_I-\eta _i^\alpha \xi ^-_I \label{apphevar1}\\
	\tilde{\lambda}_{iJ}^{\dot{\alpha}}&= -\tilde{\lambda}_i^{\dot{\alpha}} \xi ^-_J+\tilde{\eta}_i ^{\dot{\alpha}}\xi ^+_J~. \label{apphevar2}
\end{align}
Here $(I,J)$ are little group indices and $\xi ^{\pm}_I$ are two orthonormal 2-component vectors which serves as basis in the $SU(2)$ space. We choose them as follows
\begin{align}
\xi ^{-I} =(1~~0)~,\quad \xi ^{+I}=(0~~1)~.
\end{align}
The little group indices are raised or lowered with the totally antisymmetric $2\times 2$ matrix $\varepsilon _{IJ}$.
Using the expressions in \eqref{apphevar1} and \eqref{apphevar2}, momentum matrix $p_{i\alpha \dot{\alpha}} = p_{i, \mu}\sigma^{\mu}_{\alpha \dot{\alpha}}$ for the massive particle can be written as follows:
\begin{align}
	p_{i\alpha \dot{\alpha}}=-\lambda _{i\alpha}\tilde{\lambda}_{i\dot{\alpha}}+\eta _{i\alpha}\tilde{\eta}_{i\dot{\alpha}}~.
\end{align}
The Dirac equation in terms of massive spinor-helicity variables takes the following form:
\begin{align}
p_{\alpha \dot{\alpha}}\lambda ^\alpha _I=-m\tilde{\lambda}_{I\dot{\alpha}}~,\quad p_{\alpha \dot{\alpha}} \tilde{\lambda}^{\dot{\alpha}}_I=m\lambda _{I\alpha}~. \label{dirac}
\end{align} 
Using the above equation one can show that
\begin{align}
p_{i\mu}e^{\mu}_{IJ}(i)=0~.
\end{align}
We adopt the same momentum representation as in \citep{Arkani-Hamed:2017jhn} for the $(\lambda _\alpha, \tilde{\lambda}_{\dot\alpha})$, $(\eta _\alpha, \tilde{\eta}_{\dot\alpha})$ variables so that they satisfy $\langle \lambda _i \eta _i \rangle =[\tilde{\lambda}_i\tilde{\eta}_i]=m$. We use these relations to replace the mass present in the denominator in \eqref{massivepol1}. The polarization tensor for massive particle  can be now expressed in terms of massless spinors as
\begin{align}
	e^\mu _{IJ}(i)=\left[\frac{\langle \lambda _i |\sigma ^\mu |\tilde{\eta}_i]}{\sqrt{2}[\tilde{\lambda}_i\tilde{\eta}_i]}\xi ^+_I \xi ^+_J -\frac{\langle \eta _i|\sigma ^\mu |\tilde{\lambda}_i]}{\sqrt{2}\langle \eta _i \lambda _i  \rangle}\xi ^-_I \xi ^-_J    \right] -\frac{1}{2\sqrt{2}m}\left( \langle \lambda _i |\sigma ^\mu |\tilde{\lambda}_i]+\langle \eta _i |\sigma ^\mu |\tilde{\eta}_i]\right)\xi ^+_{(I}\xi ^-_{J)}~.
\end{align}
We can now read off the transverse and longitudinal polarization vectors from the coefficients of the $\xi_I$-bilinears. The polarization vectors associated to the transverse modes are given by:
\begin{align}
	e^\mu _+(i)=\frac{\langle \eta _i|\sigma ^\mu |\tilde{\lambda}_i]}{\sqrt{2}\langle \eta _i \lambda _i  \rangle} ~, \qquad e^\mu _-(i)=\frac{\langle \lambda _i |\sigma ^\mu |\tilde{\eta}_i]}{\sqrt{2}[\tilde{\lambda}_i\tilde{\eta}_i]} \label{app:masspol}
\end{align} 
The polarization vector associated to the longitudinal mode is given by:
\begin{align}
	e^\mu _0(i)=\frac{1}{2\sqrt{2}m}\left( \langle \lambda _i |\sigma ^\mu |\tilde{\lambda}_i]+\langle \eta _i |\sigma ^\mu |\tilde{\eta}_i]\right)~. \label{longitudinal}
\end{align}

\subsection{Deformed massive polarization vector at large $z$} \label{appendix:b}
The polarization tensor for a massive particle, in terms of massive spinor-helicity variables, is given in \eqref{massivepol1}. An important component in our analysis of the two-line shifts is the behaviour of the polarization tensor at large $z$. In this limit, the transverse and longitudinal modes get decoupled from each other. 
From \eqref{app:masspol} the massless polarization vectors are function of massless spinors $(\lambda _i,\eta _i)$ and $(\tilde{\lambda}_i,\tilde{\eta}_i)$. In the large-$z$ limit we treat $(\eta _i^\alpha, \tilde{\eta}_i^{\dot{\alpha}})$ as the usual reference spinors $(\zeta ^\alpha ,\xi ^{\dot{\alpha}})$ that appear in the massless case (and which are chosen to be some of the external momenta in a given scattering amplitude calculation). So the expressions for the deformed polarization vectors, which we denote as $\widehat{e}^+_\mu(i)$ can be written as follows:
\begin{align}
\widehat{e}^+_\mu(i)=\frac{\langle \zeta |\sigma _\mu|\widehat{\tilde{\lambda}}_i]}{\sqrt{2}\langle \zeta \lambda _i\rangle}~, \quad \widehat{e}^-_\mu(i)=\frac{\langle \lambda _i|\sigma _\mu |\xi]}{\sqrt{2}[\widehat{\tilde{\lambda}}_i\xi]}~.
\end{align}
The shift for the massless spinor-helicity variable $\tilde{\lambda}_{i\dot{\alpha}}$ in \eqref{apphevar2} can be obtained from the shift of massive spinor-helicity variable $\widehat{\tilde{\lambda}}_{i\dot{\alpha}}^I$ as
\begin{align}
\widehat{\tilde{\lambda}}_{i\dot{\alpha}} &=\tilde{\lambda}_{i\dot{\alpha}}-\frac{z}{m}\tilde{\lambda}_{j\dot{\alpha}}[\tilde{\lambda}_i\tilde{\lambda}_j]. \label{lambdadef} 
\end{align}
By choosing reference spinors $\zeta ^\alpha \equiv \lambda _j ^\alpha $ and $\xi _{\dot{\alpha}}\equiv \tilde{\lambda}_{j\dot{\alpha}} $, we can express 
\begin{align}
\label{apptransvsersepol}
\widehat{e}^+_b(i)= \Sigma _{ijb}-\frac{z}{m}p_{jb}\,\Omega_{ij}~,\quad \widehat{e}^-_b(i)= \frac{\Sigma _{ijb}^*}{\sqrt{2}[\tilde{\lambda}_i\tilde{\lambda}_j]}~.
\end{align}
where we define 
\begin{align}
\Sigma _{ijb}=\frac{\langle \lambda _j|\sigma _b|\tilde{\lambda}_i]}{\sqrt{2}\langle \lambda _j \lambda _i\rangle} \quad\text{and}\quad \Omega_{ij}=\frac{[\tilde{\lambda}_i \tilde{\lambda}_j]}{\sqrt{2}\langle \lambda _j\lambda _i \rangle}~.
\end{align}
These relations have been used in \eqref{plusmap} and \eqref{minusmap}. 

\subsection{Deformed massless polarization vector}\label{appendix:c}
In order to derive the limiting behaviour of $\widehat{\mathcal{A}}^h_{IJ}$ at large $z$, we  expressed the shifted massless polarization $\widehat{e}^\mu_+(j)$ in terms of the shift momentum $r^\mu$. In this appendix we show how to derive \eqref{map}. We start with the massless polarization vector 
\begin{align}
\widehat{e}^\mu _+(j) =\frac{\langle \xi |\sigma ^\mu |\widehat{\tilde{\lambda}}_j]}{\sqrt{2}\langle \xi \widehat{{\lambda}}_j\rangle}=\frac{\langle \xi |\sigma ^\mu |\tilde{\lambda}_j]}{\sqrt{2}\langle \xi \widehat{{\lambda}}_j\rangle}~. \label{masslesspol}
\end{align}
Here $\xi _\alpha$ is a reference spinor, which we choose to be  
\begin{align}
	\xi ^\alpha =\frac{p_i^{ \dot{\beta}\alpha}}{m} \tilde{\lambda}_{j\dot{\beta}}~.
\end{align}
Then the massless polarization vector can be written as 
\begin{align}
	\widehat{e}^\mu _+(j) =\frac{1}{\sqrt{2}\langle \lambda _j |p_i|\tilde{\lambda}_j]} p_i^{\dot{\beta}\alpha}\tilde{\lambda}_{j\dot{\beta}}\tilde{\lambda}_j^{\dot{\alpha}} \sigma ^\mu _{\alpha \dot{\alpha}}~.
\end{align}
Recall the expression for the shift momentum $r^\mu$ in terms of spinor-helicity variables (see equation \eqref{refrdef}):
\begin{align}
	r^\mu =-\frac{1}{2m}\bar{\sigma}^{\mu \dot{\alpha}\alpha}p_{i\alpha \dot{\beta}}\tilde{\lambda}_{j\dot{\alpha}}\tilde{\lambda}_j^{\dot{\beta}}=\frac{1}{2m}p_i^{\dot{\beta}\alpha}\tilde{\lambda}_{j\dot{\beta}}\tilde{\lambda}_j^{\dot{\alpha}} \sigma ^\mu _{\alpha \dot{\alpha}}~.
\end{align}
Thus we can write massless polarization vector as follows
\begin{align}
	\widehat{e}^\mu _+(j) =\frac{\sqrt{2}m r^\mu}{\langle \lambda _j |p_i|\tilde{\lambda}_j]}\equiv\kappa\, r^\mu ~.
\end{align}
We have used this result in \eqref{map}.

\section{Examples of Invalid Shifts}\label{appendix:B}

In this appendix we show that the massive-massless shifts $[\textbf{m}-\rangle $ and $[+\textbf{m} \rangle$ are invalid as the deformed amplitude does not vanish at large $z$. Let us first consider the $[\textbf{m}-\rangle $ shift. The massless polarization $\widehat{e}^-_b(j)$ takes the form:
\begin{align}
\widehat{e}^-_b(j) =\frac{\langle \widehat{j}|\sigma _b|\eta]}{\sqrt{2}[j \eta]}=\frac{\langle j |\sigma _b|\eta ] +\frac{z}{m}[i_Ij]\langle i^I|\sigma _b|\eta ]}{\sqrt{2}[j \eta]}~. \label{negetivehelicity}
\end{align} 
In the last step we have used the shift equation \eqref{masslessshift}. By choosing the reference spinor to be 
\begin{align}
\tilde{\eta}_{\dot{\alpha}} =\frac{p_{i\alpha \dot{\alpha}}}{m}\lambda _j ^\alpha~,
\end{align}
the negative helicity polarization vector in \eqref{negetivehelicity} becomes 
\begin{align}
\widehat{e}^-_b(j) =\frac{ 2mr_b^* +\frac{z}{m}\left[ (p_i \cdot p_j )p_{ib} -2p_{jb}\right]}{\sqrt{2}\langle j|p_i|j]} ~.
\end{align}
The transverse polarizations of the massive particle will be same as in \eqref{apptransvsersepol}. The natural amplitudes for the transverse modes have the following large-$z$ behaviour: 
\begin{align}
\widehat{\mathcal{A}}^-_- &=\frac{1}{\sqrt{2}\langle j|p_i|j]}\left[M^{ab} +\frac{1}{z}B^{ab}+... \right]d_{ijb} \left( 2mr_b^* +\frac{z}{m}\left[ (p_i \cdot p_j )p_{ib} -2p_{jb}\right] \right) \cr
& \sim \mathcal{O}(z) ~,\\
\text{and}\hspace{0.7cm}\widehat{\mathcal{A}}^-_+ & \sim \mathcal{O}(z^2)~,
\end{align}
thereby proving that the $[\textbf{m}-\rangle $ shift is invalid. One can similarly show that the $[+\textbf{m} \rangle$ shift is also invalid.

\end{appendix}

\bibliographystyle{JHEP}
\bibliography{self}

\end{document}